\documentclass[fleqn,10pt]{wlscirep}
\usepackage[utf8]{inputenc}
\usepackage[T1]{fontenc}
\usepackage{lineno}
\usepackage{subcaption}
\usepackage{cleveref}
\usepackage{graphicx}
\usepackage{color,soul}
\usepackage{makecell}
\usepackage{tabularx}
\usepackage{multirow}
\usepackage{rotating}
\usepackage{array}
\usepackage{longtable}
\usepackage{filecontents,catchfile}
\usepackage{longtable}
\usepackage{caption}
\usepackage{diagbox}

\title{The smarty4covid dataset and knowledge base: a framework enabling interpretable analysis of audio signals}

\author[1,*]{Konstantia Zarkogianni}
\author[1]{Edmund Dervakos}
\author[1,$\dag$]{George Filandrianos}
\author[1,$\dag$]{Theofanis Ganitidis}
\author[1]{Vasiliki Gkatzou}
\author[2]{Aikaterini Sakagianni}
\author[3]{Raghu Raghavendra}
\author[3]{C.L. Max Nikias}
\author[1]{Giorgos Stamou}
\author[1,3]{Konstantina S. Nikita}

\affil[1]{National Technical University of Athens, School of Electrical and Computer Engineering, Athens, 157 80, Greece}
\affil[2]{Sismanoglion General Hospital, Department of Intensive Care Unit, Athens, 15126, Greece}
\affil[3]{University of Southern California, Viterbi School of Engineering, Los Angeles, 90089, USA}

\affil[*]{corresponding author(s): Konstantia Zarkogianni (kzarkog@biosim.ntua.gr)}

\affil[$\dag$]{these authors contributed equally to this work}

\begin{abstract}
Harnessing the power of Artificial Intelligence (AI) and m-health towards detecting new bio-markers indicative of the onset and progress of respiratory abnormalities/conditions has greatly attracted the scientific and research interest especially during COVID-19 pandemic. The smarty4covid dataset contains audio signals of cough (4,676), regular breathing (4,665), deep breathing (4,695) and voice (4,291) as recorded by means of mobile devices following a crowd-sourcing approach. Other self reported information is also included (e.g. COVID-19 virus tests), thus providing a comprehensive dataset for the development of COVID-19 risk detection models. The smarty4covid dataset is released in the form of a web-ontology language (OWL) knowledge base enabling data consolidation from other relevant datasets, complex queries and reasoning. It has been utilized towards the development of models able to: (i) extract clinically informative respiratory indicators from regular breathing records, and (ii) identify cough, breath and voice segments in crowd-sourced audio recordings. A new framework utilizing the smarty4covid OWL knowledge base towards generating counterfactual explanations in opaque AI-based COVID-19 risk detection models is proposed and validated.      
  
\end{abstract}
\begin{document}

\flushbottom
\maketitle

\thispagestyle{empty}

\section*{Background \& Summary}
The COVID-19 pandemic induced innovation in many technological sectors leading to the development of a variety of means to combat the global outbreak such as vaccines, bio-sensors facilitating the diagnosis at the point of care, 3D printed ventilators and a wealth of mobile applications. More specifically, leveraging the latest trends on mobile health technologies, several applications have been implemented to fight COVID-19 with the aim of creating awareness, collecting suitable data for health survey and surveillance, reducing person-to-person contacts, offering telemedicine services, tracking COVID-19 contacts, supporting healthcare professionals in decision making, facilitating communication and collaboration among healthcare providers and serving as a means towards coordinating emergency response and transport \cite{williams2020mobile}. Further to the above, Artificial Intelligence (AI) and Machine Learning (ML) have played an important role in the response to the COVID-19 related challenges through accelerating the research and treatment while offering remarkable solutions to the diagnosis taking into consideration several biomedical data such as X-rays, Computer Tomography (CT) scans, electrocardiogram and audio recordings \cite{mhlanga2022role,adamidi2021artificial}. Furthermore, ML has demonstrated promising performance in epidemiological modelling based on social and weather data \cite{athanasiou2023long}.   

A prompt diagnosis of newly infected cases is of particular importance. However, RT-PCR tests and CT scans suffer from certain limitations such as variable sensitivity and increased turnaround time while requiring highly trained staff, approved laboratories and expensive equipment. The antigen tests constitute an alternative, nevertheless they demonstrate poor sensitivity \cite{Aleixandre2022}. An m-health approach able to support affordable, fast, sustainable and effective testing facilitating multiple repetitions to track progression, could contribute in containing the spread and suppressing resurgence \cite{Han2022}. Within this context, the idea of harnessing the power of AI coupled with mobile technologies to implement an easy-to-use and widely accessible COVID-19 detection method, has motivated the application of signal analysis and AI on audio recordings of cough, voice and breath towards the detection of innovative COVID-19 related bio-markers \cite{Aleixandre2022,Ghrabli2022}.          

\begin{figure}[t]
    \centering
    \includegraphics[scale=0.7]{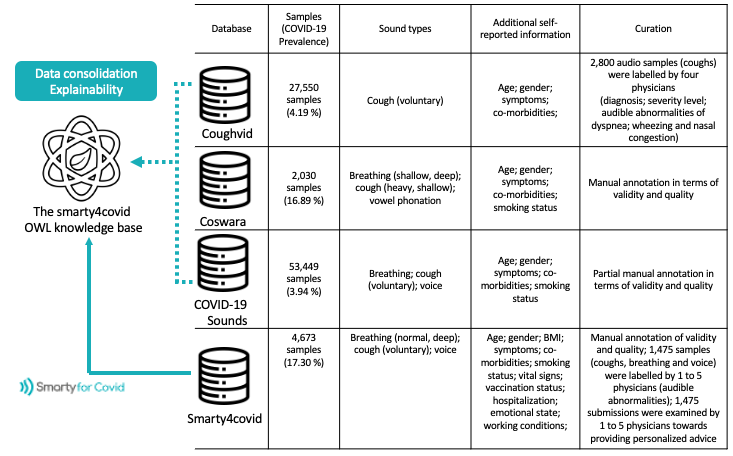}
    \caption{The smarty4covid contribution. Each sample is considered to contain audio recordings of all sound types that are collected within the frame of each crowd-sourcing approach}
    \label{fig:Infographics}
\end{figure}

In recent literature, most approaches for predicting the COVID-19 risk from audio recordings rely on deep learning models, which typically require large amounts of data to be trained. Therefore the development of curated COVID-19 audio datasets is crucial for achieving accuracy and reliability \cite{Ghrabli2022}. Several studies have been oriented to collect audio recordings from citizens following a crowd-sourcing approach through the use of a web interface. The first attempt in this direction has been initiated within the frame of the COVID-19 Sounds project \cite{xia2021covid}. The COVID-19 Sounds database consists of 53,449 audio samples each including 3 to 5 deep breaths through mouth, 3 voluntary coughs and 3 voice repetitions of a predefined short sentence. Coswara is another crowd-sourced database consisting of various kinds of sounds such as breaths (shallow and deep), voluntary coughs (heavy and shallow), sustained vowel phonation (/ey/ as in made, /i/ as in beet, /u:/ as in cool), and number counting from one to twenty (normal and fast-paced) \cite{sharma2020coswara}. Coughvid is also considered to be one of the largest crowd-sourced databases, yet including only cough sounds \cite{orlandic2021coughvid}. To date, the latest version of Coughvid is publicly released with 27,550 cough recordings. As illustrated in Figure \ref{fig:Infographics} the number of the obtained audio samples (e.g. each audio sample includes all the considered types of audio recordings) ranges from 2,030 to 53,449 while the prevalence of COVID-19 cases is relatively low especially in the Coughvid and the COVID-19 Sounds datasets. All these available databases include various demographics, symptoms, and co-morbidities in order to provide further information towards detecting COVID-19. 

A common pitfall of crowd-sourced data is that it contains audio recordings unrelated to the desired content of the database and audio recordings characterized by low quality and increased noise. This highlights the need to apply methods for data curation. Within the frame of the Coughvid and the COVID-19 Sounds projects, computational models have been developed to detect the specific segments in the audio signal that contain the considered audio recording. More specifically, the YAMNet pre-trained audio classification network has been used to filter out noisy, silent, low-quality, and inconsistent recordings \cite{xia2021covid} in the dataset. The model has been evaluated on a small subset (e.g. 3,067 audio recordings) that has been manually annotated and has achieved an accuracy up to 88\%. In the case of the Coughvid dataset, a small number (e.g 215) of audio recordings have been selected and manually annotated as cough or non-cough sounds. This small datataset has been used to develop an eXtreme Gradient Boosting classifier towards discriminating cough from non cough audio recordings taking as input 68 audio features in the domains of (i) Mel Frequency, (ii) Time, and (iii) Frequency. Following a 10-fold cross validation framework the COUGHVID model has achieved sensitivity and c-statistic equal to 78.2\% and 96.4\%, respectively \cite{orlandic2021coughvid}. The Coswara dataset has been entirely manually annotated.          

The development of robust machine learning models able to detect COVID-19 is particularly challenging due to the heterogeneity of the available datasets, the low number of cases (positive for COVID-19) versus controls (negative for COVID-19), and deficiencies related to COVID-19 variants and factors that strongly affect the infection, for example the vaccination status against COVID-19. On top of this, there are biases in the available datasets that need to be thoroughly investigated, while there is an increased risk of model over-fitting especially when complex modelling strategies are applied \cite{Han2022}. The realistic performance of an audio based digital testing for COVID-19 has been explored through artificially creating biases in the development dataset, for example introducing gender bias into the data by selecting a high percentage of cases as males, and evaluating their impact on the model's efficacy \cite{Han2022}. Another research challenge is the development of a knowledge representation of the available data/information that enables data consolidation and reasoning. The latter is particularly important in order to ensure transparency and gain end-users' trust through providing explanations of the estimated risk. From this perspective, the deployment of smart interfaces that present end users with human understandable interpretations and explanations of their estimated COVID-19 probability can greatly support informed decision making while enhancing human supervision towards the realization of a human centered AI approach. The development of responsible AI models requires data that is richly annotated with metadata, expert labels, and semantic information. This additional information can be used as high-level features for training explainable AI models, since these features are more understandable for humans than for example audio signals or spectrograms that usually form the input space of deep learning models. Furthermore, this additional information can be utilized for post hoc explainability and analysis of black-box classifiers, which is particularly useful since opaque deep learning models are usually applied towards detecting COVID-19 from audio recordings \cite{brown2020exploring,chaudhari2020virufy,cohen2020novel}.   

The smarty4covid project aspires the creation of an intelligent multimodal framework for COVID-19 risk assessment and monitoring based on Explainable Deep Learning. Following the necessary approvals from the National Technical University’s Ethics Committee of Research, a responsive web based application (www.smarty4covid.org) has been implemented and publicly released as a means of data collection. The smarty4covid dataset contains in total 18,265 audio recordings of cough, breath (regular, deep) and voice corresponding to 4,673 users (Greek and Cypriot citizens). It also includes other self-reported information related to demographics, symptoms, underlying conditions, smoking status, vital signs, COVID-19 vaccination status, hospitalization, emotional state, working conditions and COVID-19 status (e.g. positive, negative, not-tested). The entire dataset has been cleaned of erroneous and noisy samples, and a subset of the dataset (e.g. 1,475 samples) has been labeled by medical experts. Furthermore, all available information has been encoded into an innovative web ontology knowledge (OWL) base that also contains a rudimentary hierarchy of concepts. The medically related concepts in the OWL knowledge base are provided in the form of ids from SNOMED-CT \cite{noauthor_snomed_nodate}.

The curated crowd-sourced smarty4covid dataset is publicly released, yet all audio records of voices that are considered personal data according to the GDPR regulation are excluded (Figure \ref{fig:Infographics}). The smarty4covid OWL knowledge is also made available in order to enable data consolidation from multiple databases. The smarty4covid OWL knowledge base offers an interpretable framework of high expressiveness which can be employed to explain complex machine learning models through identifying semantic queries over the knowledge that mimic the model \cite{liartis2021semantic}. The smarty4covid dataset has been utilized towards the development of models able to: (i) classify segments of audio signals as "cough", "breath", "voice", and "other", and (ii) detect inhalation and exhalation segments from breathing recordings, that can be used for extracting clinically related features such as respiratory rates (RR), inhalation to exhalation ratio (I/E ratio), and fractional inspiration time (FIT). The smarty4covid OWL knowledge has been validated as a means of generating counterfactual explanations and discovering potential biases in the available datasets.   
 
\section*{Methods}
\begin{figure}[h!]
    \centering
    \includegraphics[scale=0.5]{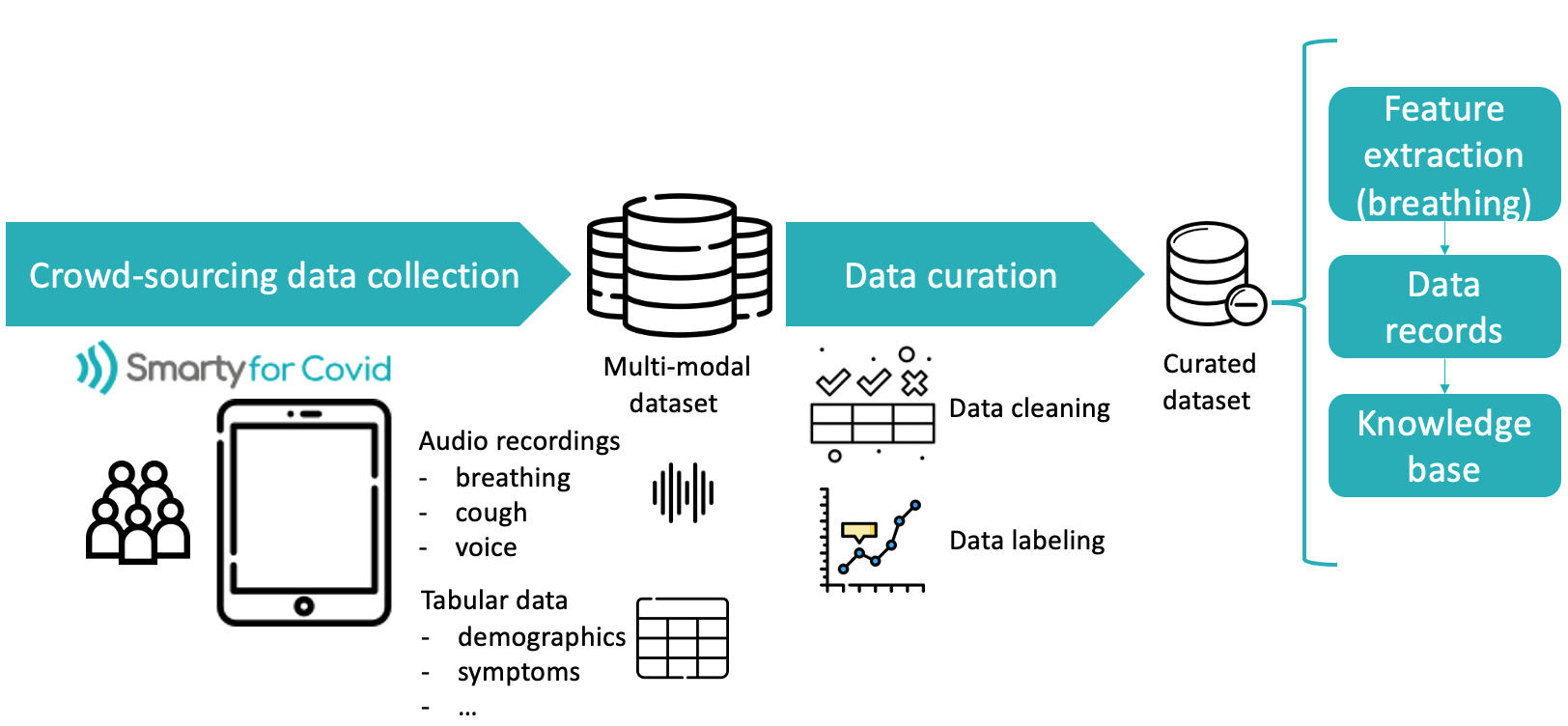}
    \caption{Overall approach towards developing the smarty4covid database}
    \label{fig:Overall}
\end{figure}

The overall approach towards the development of the smarty4covid database is depicted in Figure \ref{fig:Overall}. It includes a crowd-sourcing data collection strategy followed by a two-step data curation method involving data cleaning and labeling. A multi-modal dataset was collected including audio records and tabular data. The curated dataset was exploited for extracting breathing related features, creating publicly available data records, and developing the smarty4covid OWL knowledge that enables data selection and reasoning.       

\subsection*{Crowd-sourcing Data Collection}

The smarty4covid crowd-sourcing data collection was approved by the National Technical University’s Ethics Committee of Research and complied with all relevant ethical regulations. A responsive and user-friendly web-based application (www.smarty4covid.org) was implemented targeting Greek and Cypriot citizens older than 18 years old. The smarty4covid questionnaire consisted of several sections accompanied by instructions for users to perform audio recordings of voice, breath and cough and provide information regarding demographics, COVID-19 vaccination status, medical history, vital signs as measured by means of oximeter and blood pressure monitor, COVID-19 symptoms, smoking habits, hospitalization, emotional state and working conditions. Four types of audio recordings were considered: (i) three voice recordings where the user was required to read a specific sentence, (ii) five deep breaths, (iii) 30 \emph{s} regular breathing close to the microphone of the device and (iv) three voluntary coughs. Following an effective media planning, more than 10,000 individuals provided demographic information and underlying medical conditions to the smarty4covid application, yet almost half of them (e.g. 4,679) gave the necessary permissions to perform the audio recordings. The web-based application was released in January 2022 during the spread of the omicron wave in Greece, resulting in high COVID-19 prevalence (17.3\% of users were tested positive for COVID-19).        

\subsection*{Data Curation}

Part of the crowd-sourced dataset was invalid due to erroneous audio recording submissions by the users and the presence of distortions and high background noise. The data cleaning process was performed by means of a crowd-sourcing campaign utilizing the Label Studio\footnote{https://labelstud.io/} open source data labeling tool. AI engineers who volunteered to annotate the audio signals, signed a Non Disclosure Agreement (NDA) and granted with the necessary access permissions. A user-friendly environment was implemented enabling the annotators to listen the audio signals and answer to questions regarding their validity (yes/no) and their quality (Good, Acceptable, Poor) in terms of background noise and distortion.
In order to evaluate the quality of the annotations, a set of randomly selected audio files (e.g. 1,389) was considered more than once and up to 5 times in the annotation procedure. A high level of consistency (92.5\%) among the annotators was observed indicating that there was no need to have multiple annotators for each audio recording. 

\begin{figure}[h!]
    \centering
    \includegraphics[scale=0.5]{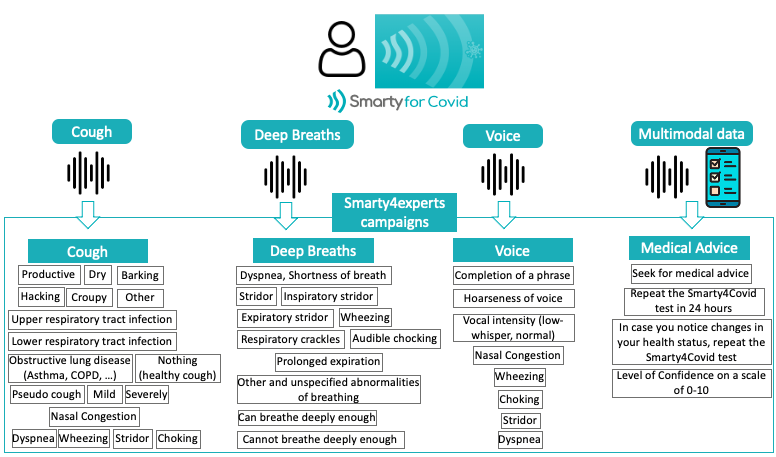}
    \caption{The smarty4covid labeling campaigns. The cough, breath, and voice campaigns include labels that are indicative of respiratory abnormalities}
    \label{fig:doc_campaign}
\end{figure}

The smarty4covid crowd-sourced dataset was enriched with labels annotated from healthcare professionals (e.g. pulmonologists, anesthesiologists, internists) who volunteered to characterize the collected audio recordings in terms of audible abnormalities and to provide personalized recommendations regarding the need for medical advice. To this end, four crowd-sourcing campaigns were initiated utilizing the Label Studio. Three campaigns focused on the audio recordings (e.g. breath, voice, cough). As depicted in Figure \ref{fig:doc_campaign}, the healthcare professionals were asked to assess the presence of audible abnormalities by selecting one or more options from the available labels. In the fourth campaign, the healthcare professionals were exposed to all available multimodal information about the user, excluding vital signs (e.g. oxygen saturation, beats per minute (BPM), diastolic/systolic pressure) that would lead them to a biased assessment, in order to estimate the risk of health deterioration and suggest a next course of action: a) Seek for medical advice, b) Repeat the Smarty4Covid test in 24 hours and c) In case you notice changes in your health status, repeat the Smarty4Covid test. They were also asked to define a level of confidence (from 1 to 10) in their assessment.  

\subsection*{Breathing Feature Extraction}
Respiration is a complex physiological process, involving both voluntary and involuntary processes, as well as underlying reflexes. A breathing pattern is the upshot of a fine coordination between peripheral chemoreceptors, central nervous system’s organizing structures, lung mechanoreceptors and parenchyma, musculoskeletal components, intrinsic metabolic rate, emotional state, and many others. A breathing pattern adopted at any given moment is assumed to be that which produces adequate alveolar ventilation at the lowest possible energy cost, given the contemporary system’s mechanical status and organism’s metabolic needs. Any disruption in any of these respiratory homeostasis’ pillars, will be reflected in a change of the respiratory pattern, shifting this balance to the best for the prevailing conditions energetic state \cite{mortola2019breathe}. A viral infection could be a breathing pattern’s disorientation factor \cite{giannakopoulou2019regulation,van2022persistent, higenbottam1982glottis, chang2004perceived}. Some quantitative indicators commonly used to describe a breathing pattern and its readjustments are RR, respiratory phases and volumes, gases partial pressure, blood gases analysis and other  \cite{mortola2019breathe,tobin1992breathing}. 

Most of the studies associated with COVID-19 crowd-sourced databases of breathing audio recordings explore features generated through signal processing or deep learning. The smarty4covid dataset innovates the current state of the art by including clinically relevant important and informative respiratory indicators extracted from regular breathing records, such as the RR, I/E ratio, and FIT. RR is the number of breaths per minute, that is normally 16-20 breaths/\textit{min.} It can be affected by both external and internal factors such as the temperature, endogenous acid-base balance, metabolic state, diseases, injuries, toxicity, etc. I/E ratio is the ratio between the inspiratory (\textit{$T_{i}$}) and expiratory time (\textit{$T_{e}$}) and it can be indicative to a flow disturbance in the respiratory tract \cite{shakhih2019assessment}. Normal breathing usually presents 1:2 or 1:3 I/E ratio at rest \cite{shakhih2019assessment} while airways obstruction may lead to prolonged expiration or inspiration resulting to an abnormal I/E ratio. FIT, also termed as the inspiratory "duty cycle" of the respiratory system, is the ratio between (\textit{$T_{i}$}) and the duration of a total respiratory cycle (\textit{$T_{tot}$}) \cite{tobin1992breathing}. It provides a rough measure of airway obstruction and stress on the respiratory muscles. Table \ref{tab:normal_breath} summarizes the description and the normal ranges of the aforementioned respiratory indicators.

A two step approach was developed in order to extract \textit{$T_{i}$} and \textit{$T_{e}$} from the crowd-sourced breathing audio signals: (i) localization of the segments on the audio signal that contains breathing, and (ii) detection of the exhaling and inhaling parts. In the first step, an AI-based model, described in the "Technical Validation" Section, was applied. The obtained breathing segments were split into non silent intervals. The second step was particularly challenging since either the inhalation part, that was characterized by low mean amplitude, was not appropriately captured due to the hardware of the recording device or due to the short distance of the sound source from the microphone during the exhalation phase, resulting in distortion of the waveform. In order to face this challenge, an unsupervised method was developed with the aim to identify similar parts on a single breathing audio signal that in turn could be considered as either inhalation or exhalation. This particular method presents several advantages over the state of the art \cite{frey2007clustering}, since it doesn't require a dataset of human-labeled data for training while there is no need to take into consideration prior knowledge that inhalation follows exhalation and vice versa. Furthermore, the application of the unsupervised method on a single breathing audio signal adds robustness against distortion and background noise since all inhalation/exhalation parts of the same breathing recording are subject to the same level of distortion and background noise.

The unsupervised method featured a clustering algorithm based on affinity propagation \cite{frey2007clustering} at a frequency level. To this end, the mel-spectrogram (MFCC-128) of the audio signal was obtained and transformed into a vector of 128 frequencies each one corresponding to the summation of the respective frequencies over time. The obtained clusters were labeled as "inhalation", "exhalation" or "other" though applying a heuristic approach. More specifically, for each cluster, the mean amplitudes were calculated by averaging the mean amplitudes over all the members of the cluster. Next, the clusters were sorted from largest to smallest mean amplitude. The top listed cluster was considered as exhalation while the second cluster (if existed) as inhalation. The remaining clusters were labeled as "other".  

For validation purposes, the inhalation and exhalation parts of thirty-three audio recordings of regular breathing, were manually annotated in order to enable the calculation of the corresponding respiratory indicators. The proposed unsupervised method achieved Root Mean Square Error (RMSE) up to 1.85, 0.14, and 0.08 for the RR, FIT, and I/E ratio, respectively. These RMSE values are considered to be low taking into consideration the normal ranges of each respiratory indicator (\ref{tab:breathing_features}). 

\section*{Data Records}
\begin{figure}
    \centering
    \includegraphics[scale=0.4]{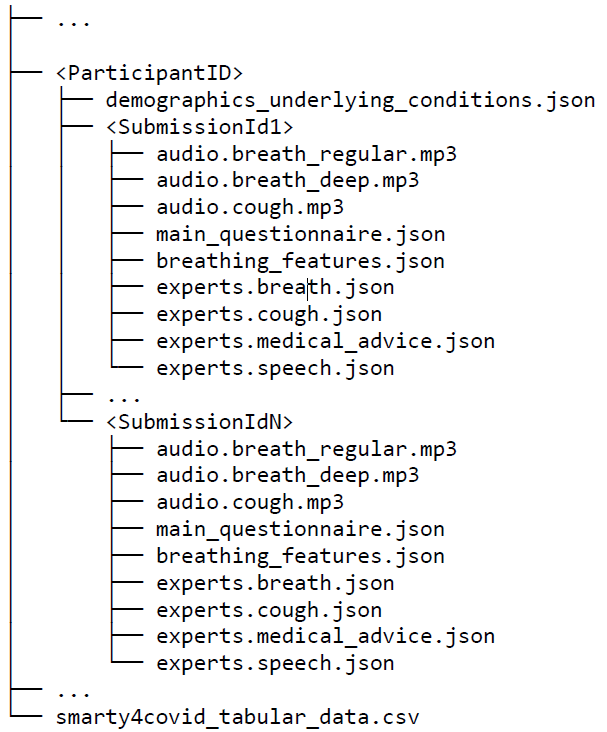}
    \caption{Structure of files in the smarty4covid dataset.}
    \label{fig:directory}
\end{figure}

Part of the smarty4covid crowd-sourced dataset (4,303 submissions) was organized into data records in order to be publicly available. The data records are deposit in the Zenodo Repository (DOI: 10.5281/zenodo.7760170). As depicted in Figure \ref{fig:directory}, each directory contains the submissions of a specific user. The user directory is named after the user’s id that is generated according to the UUID V4 protocol. Apart from the submissions, a json file (“demographics\_underlying\_conditions.json”) with information regarding demographics (e.g. BMI, age group, gender) and potential underlying conditions (Table \ref{tab:demographics_json}) is also included. Each submission corresponds to a separate sub-directory that is named after the unique submission id and it contains:

\begin{enumerate}
    \item valid audio recordings of cough (“audio.cough.mp3”), deep breathing (“audio.breath\_deep.mp3”) and regular breathing (“audio.breath\_regular.mp3”). Each audio recording has a sampling rate of 48 kHz and a bitrate of 64 kb/s.
    \item a json file (“main\_questionnaire.json”) with information related to the COVID-19 test (result, type, and date), COVID-19 vaccination status, COVID-19 related symptoms, vital signs and more (Table \ref{tab:main_questionnaire}).
    \item a json file (“breathing\_features.json”) with the extracted respiratory indicators and the manual annotations of the breathing phases (inhalation, exhalation) on the breathing audio signal (Table \ref{tab:breathing_features}). 
    \item four json files (“experts.breath.json”, “experts.cough.json”, “experts.medical\_advice.json”, “experts.speech.json”) including the input/labels (characterization, advice) from the healthcare professionals (Tables \ref{tab:experts_breathing} - \ref{tab:experts_voice}).
\end{enumerate}

\subsection*{Knowledge Base}

A web-ontology language (OWL) knowledge base \footnote{https://www.w3.org/OWL/} was developed motivated by the need of data consolidation from different relevant databases (e.g Coughvid, COVID-19 sounds, Coswara) and the application of complex queries for the detection of users with specific characteristics. All available information resulting from the crowd-sourcing, data cleaning and data labeling procedures were also released in the form of the smarty4covid OWL knowledge base. In general, using a vocabulary $\mathcal{V}=\langle\mathsf{CN,RN,IN}\rangle$ where $\mathsf{CN,RN,IN}$ are mutually disjoint sets of concept names, role names and individual names respectively, a knowledge base ($\mathcal{K}=\langle\mathcal{A,T}\rangle$) can be built through creating the Assertional Database (ABox - $\mathcal{A}$) and the Terminology Database (TBox - $\mathcal{T}$). The ABox includes assertions of the form $C(a),r(a,b)$ where $C\in\mathsf{CN}$, $r\in{\mathsf{RN}}$, and $a,b\in{\mathsf{IN}}$. The TBox is a set of terminological axioms of the form $C\sqsubseteq{D}$, where $C,D\in{\mathsf{CN}}$, $r\sqsubseteq{s}$ and  $r,s\in{\mathsf{RN}}$. Based on these axioms, the hierarchies of concepts and roles can be defined in the TBox.

\begin{figure}[]
    \centering
    \begin{subfigure}[t]{0.27\textwidth}
    \includegraphics[width=\textwidth]{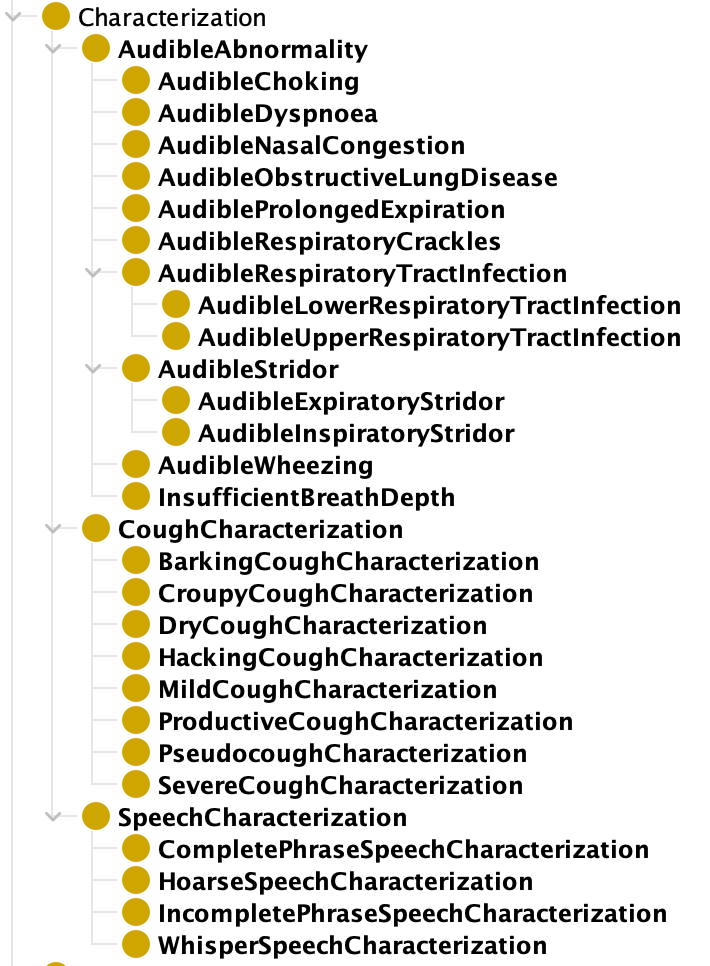}
    \caption{Concepts related to expert characterizations.}
    \label{fig:protege-audibleSymptoms}
    \end{subfigure}
    \begin{subfigure}[t]{0.27\textwidth}
    \centering
    \includegraphics[width=\textwidth]{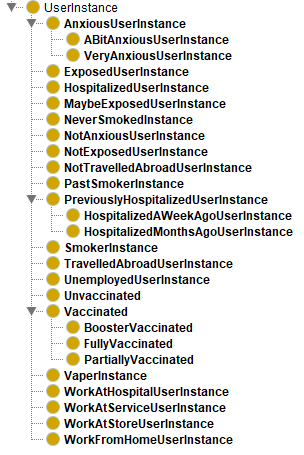}
    \caption{Questionnaire related concepts.}
    \label{fig:protege-userInstance}
\end{subfigure}
    \begin{subfigure}[t]{0.2\textwidth}
    \includegraphics[width=\textwidth]{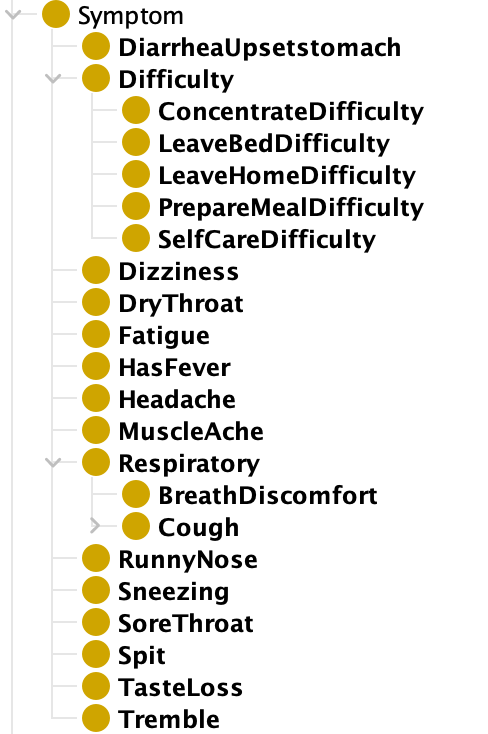}
    \caption{Symptom related concepts.}
    \label{fig:protege-symptoms}
    \end{subfigure}
    \begin{subfigure}[t]{0.2\textwidth}
    \centering
    \includegraphics[width=\textwidth]{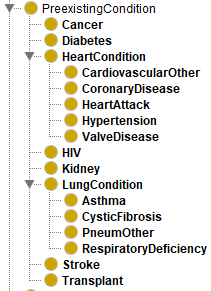}
    \caption{Preexsting conditions related concepts.}
    \label{fig:protege-preexisting}
\end{subfigure}

    \begin{subfigure}[t]{0.15\textwidth}
    \centering
    \includegraphics[width=\textwidth]{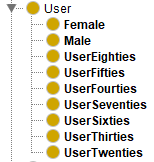}
    \caption{Concepts for the different types of users.}
    \label{fig:protege-user}
\end{subfigure}
\begin{subfigure}[t]{0.25\textwidth}
      \includegraphics[width=\textwidth]{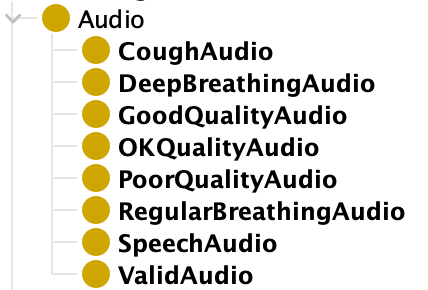}
    \caption{Audio related concepts.}
    \label{fig:protege-audio}
    \end{subfigure}
    \begin{subfigure}[t]{0.15\textwidth}
    \centering
    \includegraphics[width=\textwidth]{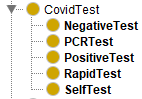}
    \caption{Concepts related to covid tests}
    \label{fig:protege-covidTest}
\end{subfigure}
    \begin{subfigure}[t]{0.25\textwidth}
    \centering
    \includegraphics[width=\textwidth]{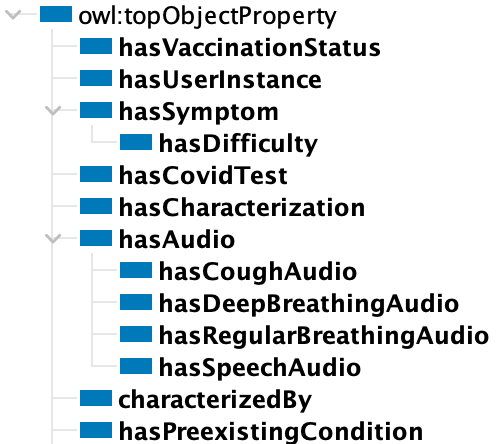}
    \caption{Hierarchy of roles.}
    \label{fig:protege-rn}
\end{subfigure}
    \caption{Hierarchies of concepts and roles from the smarty4covid knowledge base.}
    \label{fig:my_label}
\end{figure}

In the smarty4covid OWL knowledge base, the set of individual names ($\mathsf{IN}$) contains a unique name indicative to each participant, questionnaire, audio file, healthcare professional that participated in the labeling procedure and the corresponding characterizations of the audio records. ($\mathsf{IN}$) also includes unique names for each declared symptom, COVID-19 test and preexisting condition that is linked to the corresponding questionnaire (e.g symptom, COVID-19 test) and participant (e.g. underlying condition), respectively. These individuals are linked through appropriately defined roles. The role names $\mathsf{RN}$ and their defined hierarchy is depicted in Figure \ref{fig:protege-rn}. Each role is associated with a domain and a range indicative to the types of the individuals that can be linked through this role. In particular, the role $\mathsf{hasCharacterization}$ links audio files to characterizations as labelled by the healthcare professionals, and $\mathsf{characterizedBy}$ links characterizations to instances of the healthcare professionals. The role $\mathsf{hasAudio}$ and its children link questionnaires to audio files. The roles $\mathsf{hasCovidTest}$ and $\mathsf{hasSymptom}$  link questionnaires to instances of COVID-19 tests, self-reported symptoms, and vaccination status, respectively. The role $\mathsf{hasPreexistingCondition}$ links participants to preexisting conditions, while $\mathsf{hasUserInstance}$ links participants to their submitted questionnaires.

The set of concept names $\mathsf{CN}$ involves concepts that describe instances of audio, COVID-19 tests, preexisting conditions, symptoms, users and questionnaires. For audio related concepts, their hierarchy is shown is Figure \ref{fig:protege-audio}. Specifically, there is a concept for each type of audio recording (e.g. regular breathing, deep breathing, voice, cough), and concepts regarding the audio quality. Audio instances can additionally be linked, via the $\mathsf{hasCharacterization}$ role to audible abnormalities, for which the hierarchy of concepts is shown in Figure \ref{fig:protege-audibleSymptoms}. Similarly, all preexisting conditions that appear in the questionnaire are organized as concepts in a hierarchy as shown in Figure \ref{fig:protege-preexisting}, and all symptoms are part of the symptom hierarchy, shown in Figure \ref{fig:protege-symptoms}. Furthermore, the $\mathsf{User}$ concept subsumes concepts related to the different age and gender of the participants, as shown in Figure \ref{fig:protege-user}, while the $\mathsf{UserInstance}$ concept that corresponds to a specific questionnaire submitted by a user, also subsumes a hierarchy based on the different possible answers in the questionnaire, shown in Figure \ref{fig:protege-userInstance}. Finally, the concepts related to COVID-19 tests, shown in Figure \ref{fig:protege-covidTest}, are used to define the type of test and its outcome.

\begin{figure}
    \centering
    \includegraphics[width=0.8\textwidth]{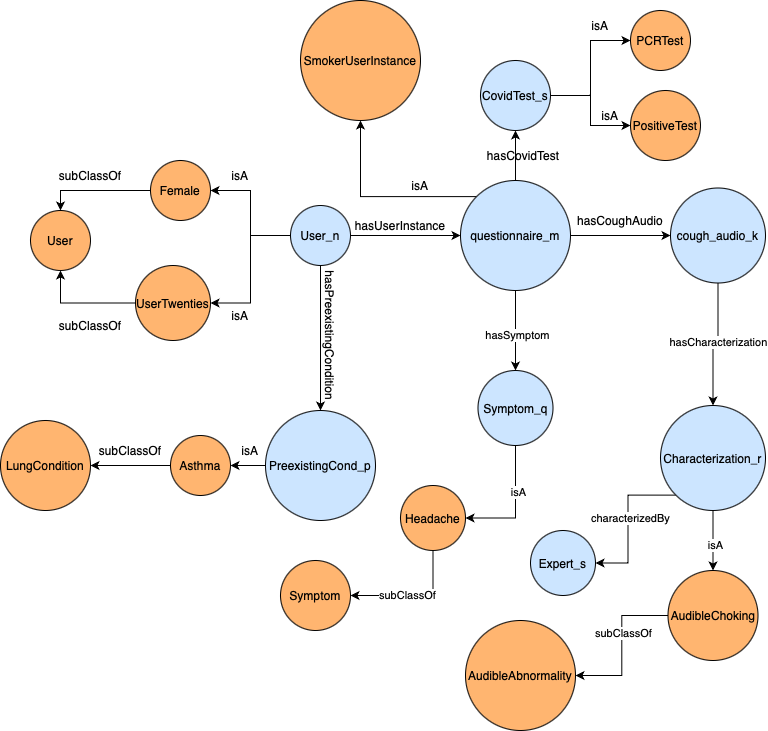}
    \caption{Example of the structure of the smarty4covid knowledge base. Blue nodes represent individuals, and orange nodes concepts. Edges labeled as IsA represent concept assertions from the ABox, and subClassOf edges represent inclusion axioms from the TBox.}
    \label{fig:knowledge_example}
\end{figure}

The described hierarchies of concepts and roles are provided in OWL format in the file [smarty-ontology.owl]. Using this terminology, all information presented in the dataset is asserted in the form of triples, provided in the file [smarty-triples.nt]. An example of a smarty4covid user is depicted in Figure \ref{fig:knowledge_example}. This user who is a female (20-30 years old) and has asthma, has submitted a questionnaire declaring a positive PCR test and a headache while being a smoker. Her audio recording of cough has been labeled by medical professionals as featuring audible choking.

\section*{Technical Validation} 
\subsection*{Inferences from statistical analysis} 
\begin{figure}[h!]
    \centering
      \begin{subfigure}{0.4\textwidth}
        \includegraphics[width=\textwidth]{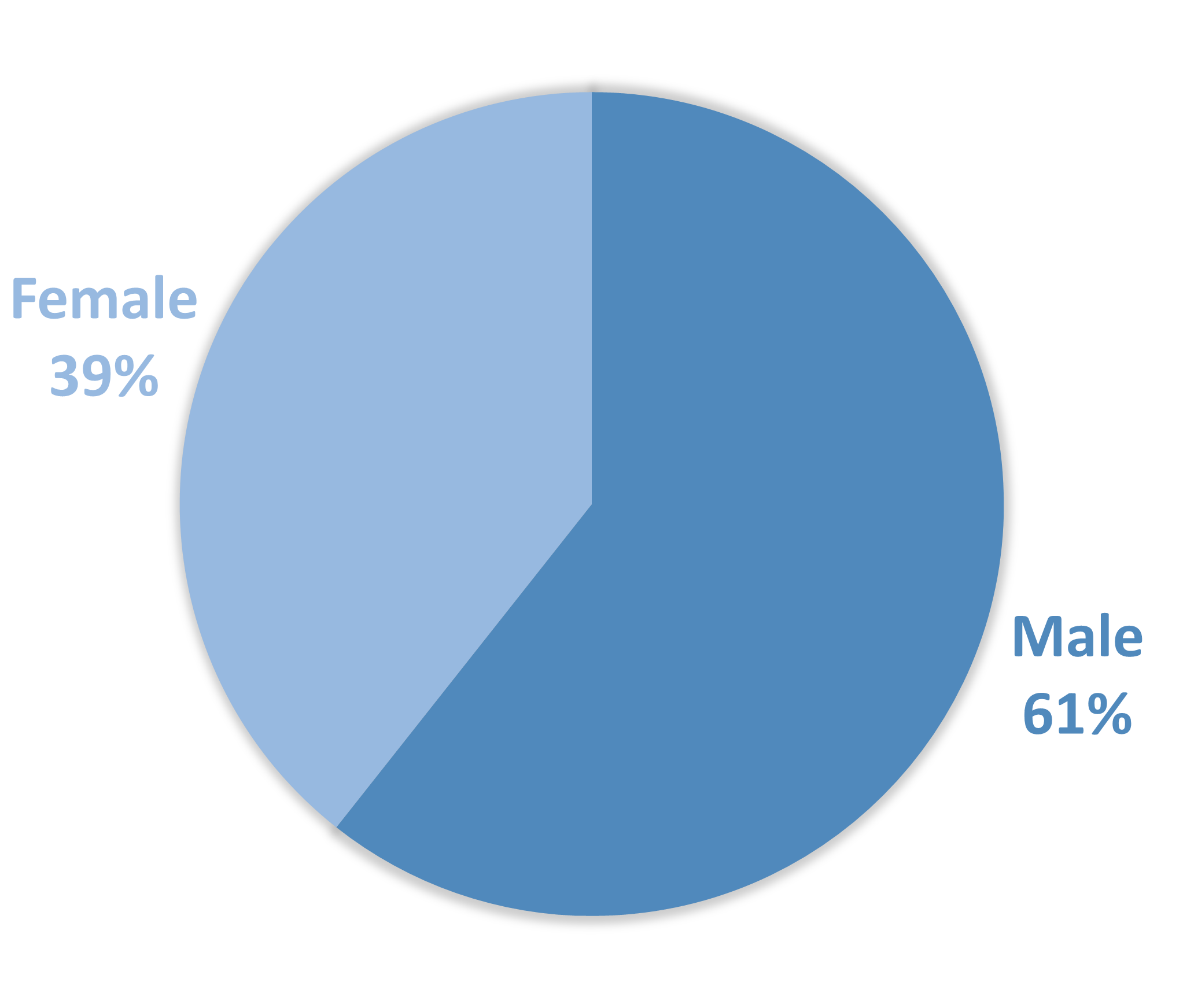}
          \caption{Sex}
          \label{fig:sex_age_results_a}
      \end{subfigure}
      \hfill
      \begin{subfigure}{0.4\textwidth}
        \includegraphics[width=\textwidth]{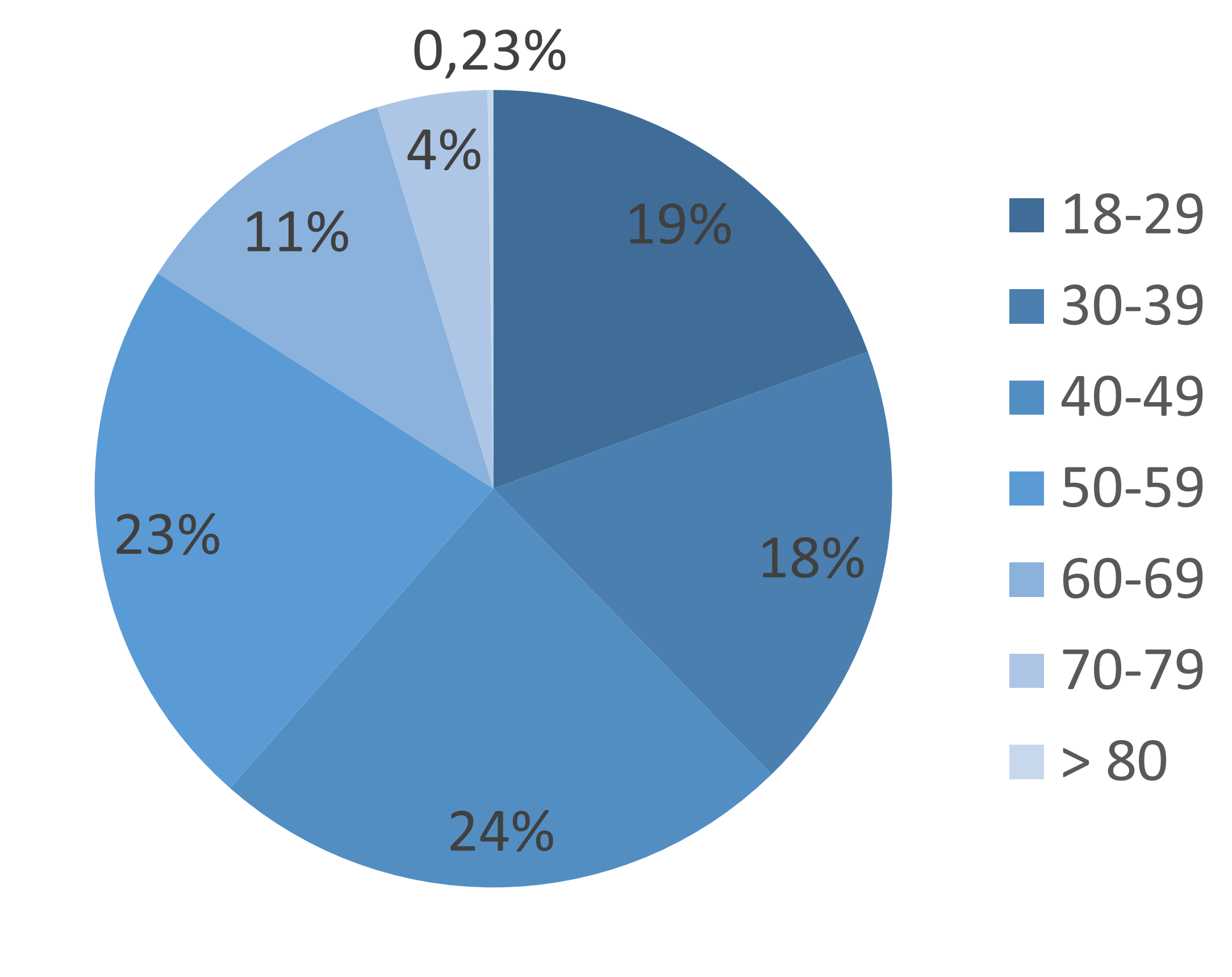}
          \caption{Age groups}
          \label{fig:sex_age_results_b}
      \end{subfigure}
      \\
      \begin{subfigure}{0.5\textwidth}
        \includegraphics[width=\textwidth]{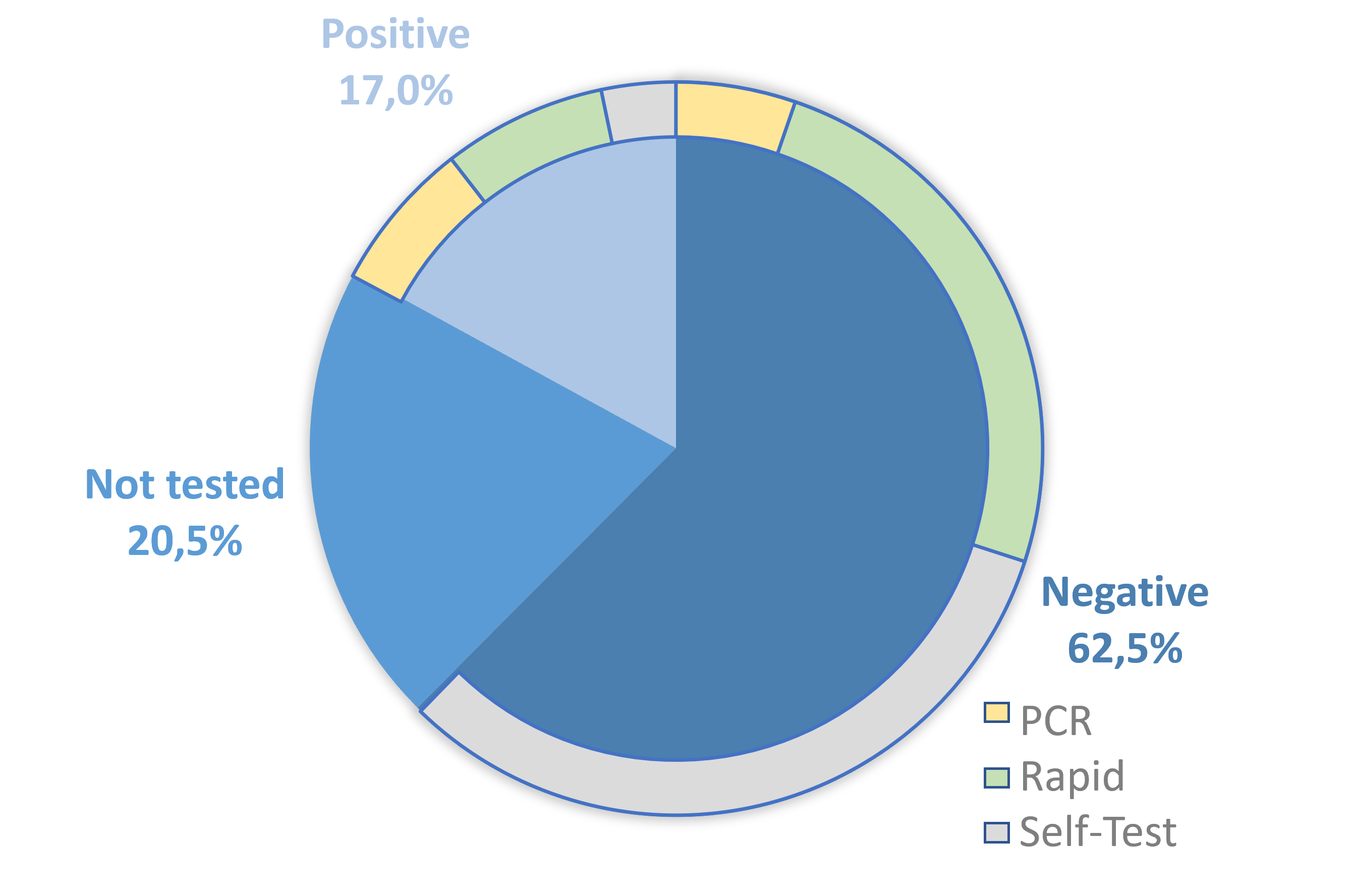}
          \caption{Test results \& types}
          \label{fig:sex_age_results_c}
      \end{subfigure}
\caption{
\label{fig:sex_age_results}%
Distribution of basic demographics, e.g. (a) gender, (b) age, and (c) COVID-19 test results.}
\end{figure}

The representativeness in the smarty4covid dataset was explored in terms of demographics, symptoms, vaccination status, COVID-19 prevalence and level of anxiety. The distribution of gender, age and COVID-19 test results is depicted in Figure \ref{fig:sex_age_results}. A higher percentage ($61.0\%$) of male versus female users was observed, yet a wide range of ages was present. Most of the users' ages were between 30 to 59 years old, that is the age range characterized by increased familiarization with mobile devices. A high percentage of submissions ($79.5\%$) included COVID-19 test results from various COVID-19 test types (e.g. PCR, Rapid Antigen, Rapid Antigen self-test) as depicted in Figure \ref{fig:sex_age_results_c}.

\begin{figure}[h!]
    \centering
    \includegraphics[scale=0.6]{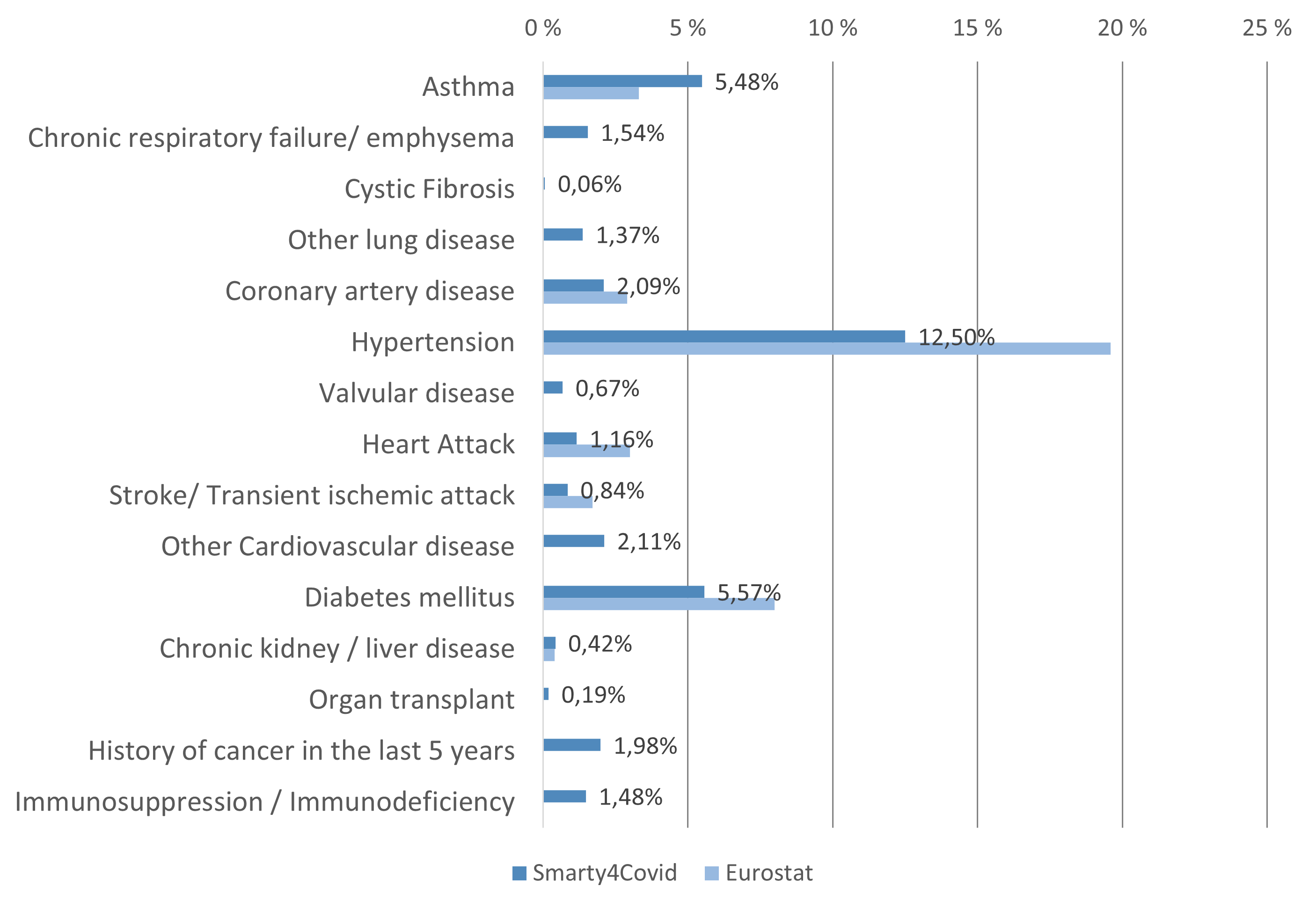}
    \caption{Distribution of underlying medical conditions.}
    \label{fig:pre_exist}
\end{figure}

Figure \ref{fig:pre_exist} illustrates the presence of underlying medical conditions associated with the progress of COVID-19 in the smarty4covid dataset. More than 1 out of 4 users ($27\%$) reported at least one underlying medical condition while hypertension  was the most commonly reported condition (Figure \ref{fig:pre_exist}). The distribution of the underlying medical conditions was similar to the one published by Eurostat \cite{eurostat} that considered the general population in Greece. 

\begin{figure}[h!]
\centering
    \begin{subfigure}{0.49\textwidth}
      \includegraphics[width=\textwidth]{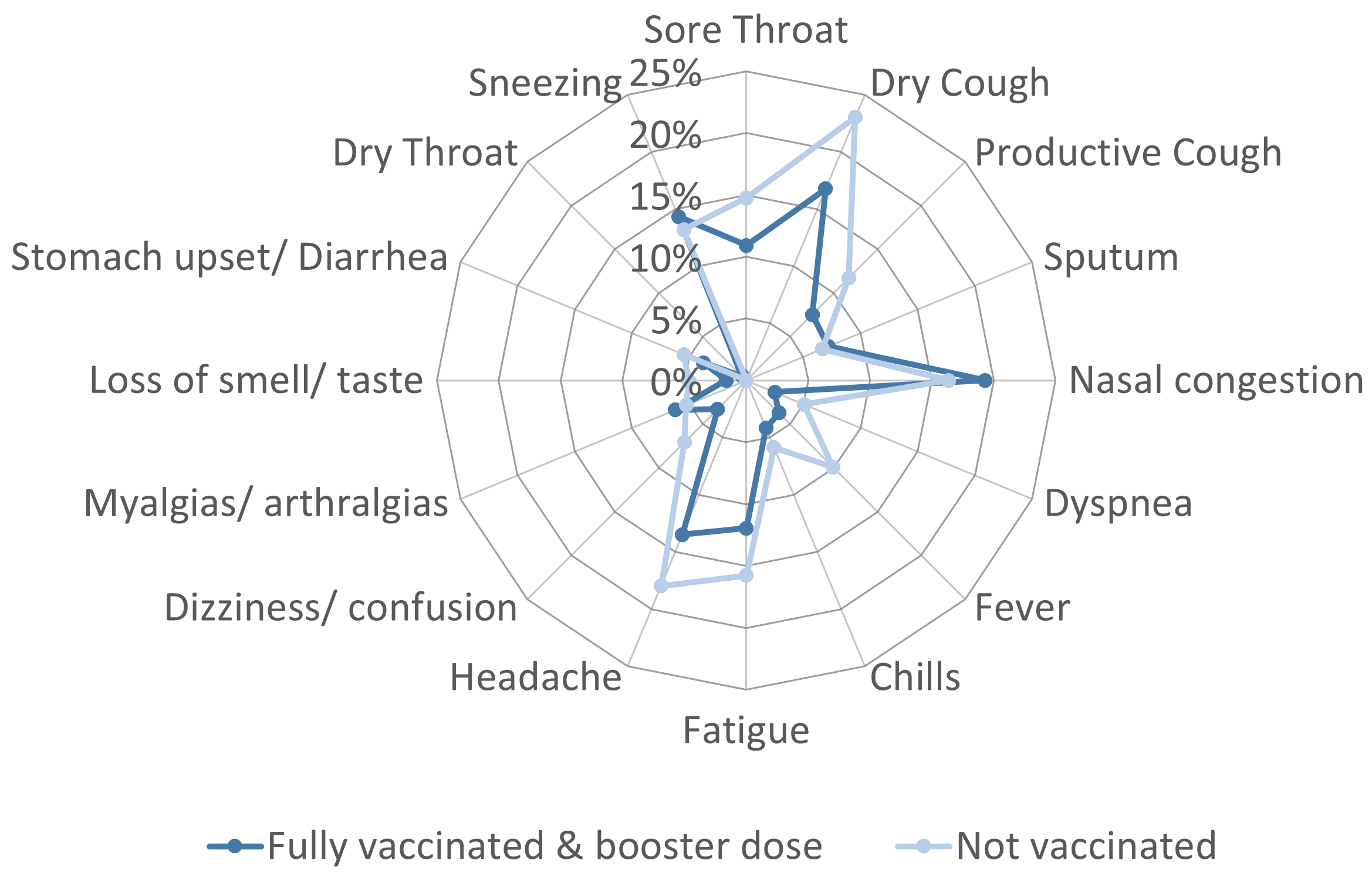}
        \label{fig:vaccination_a}%
    \end{subfigure}
    \hfill
\begin{subfigure}{0.49\textwidth}
\includegraphics[width=\textwidth]{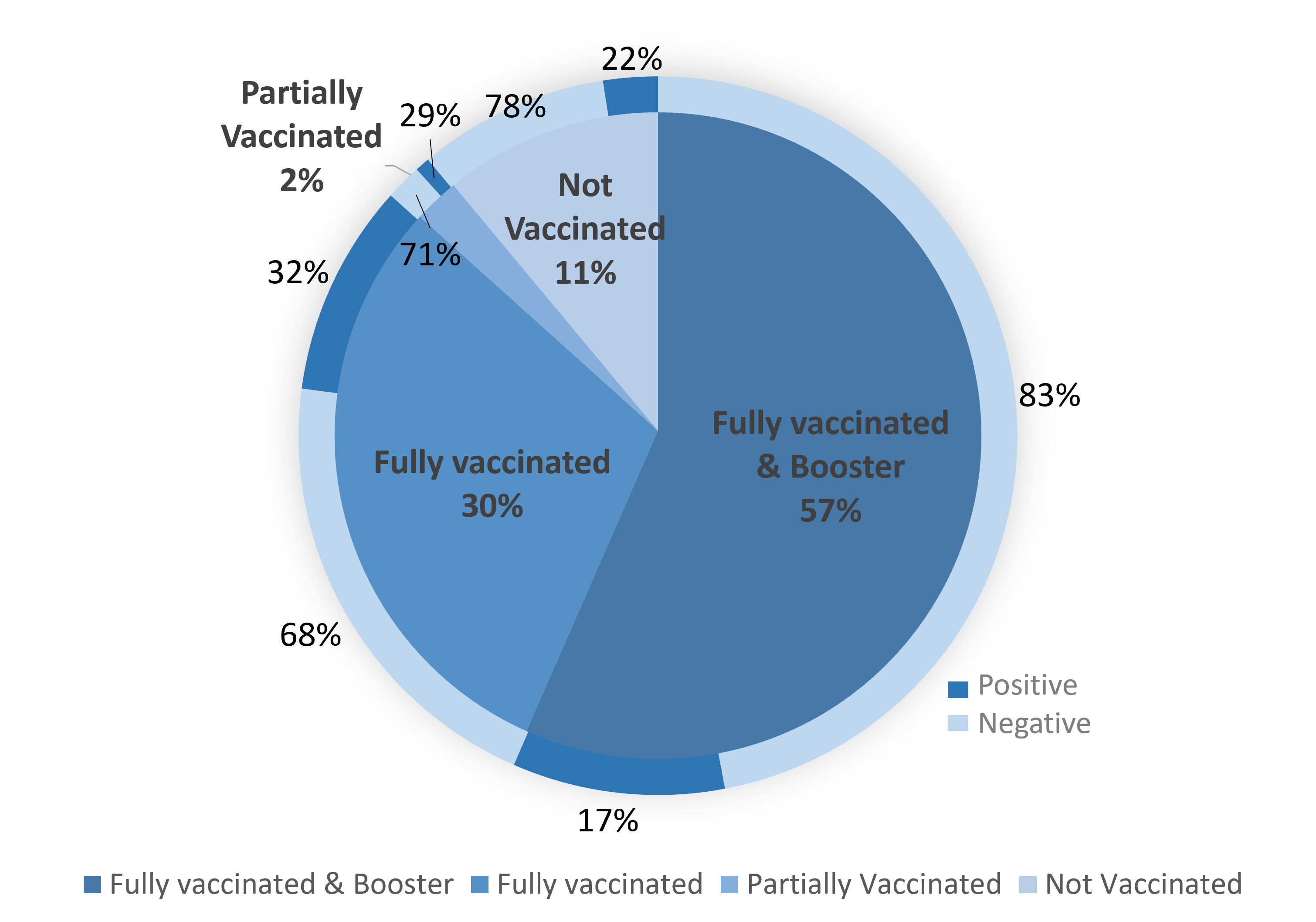}
    \label{fig:vaccination_b}
\end{subfigure}
\caption{(a) Frequency of reported symptoms, (b) Distribution of COVID-19 vaccination status and COVID-19 test results}
\label{fig:vaccination}
\end{figure}

Referring to the COVID-19 related symptoms, more than half of the users reported at least one symptom. Figure \ref{fig:vaccination} depicts the frequency of each symptom versus vaccination status (not vaccinated, fully vaccinated and booster dose). It can be inferred that users with booster dose presented fewer symptoms than those who were not vaccinated. Figure \ref{fig:vaccination} illustrates the percentages of positive and negative for COVID-19 users for each vaccination status. It can be seen that the COVID-19 prevalence is lower within the booster dose vaccinated population. 

\begin{figure}[h!]
    \centering
      \begin{subfigure}{0.49\textwidth}
        \includegraphics[width=\textwidth]{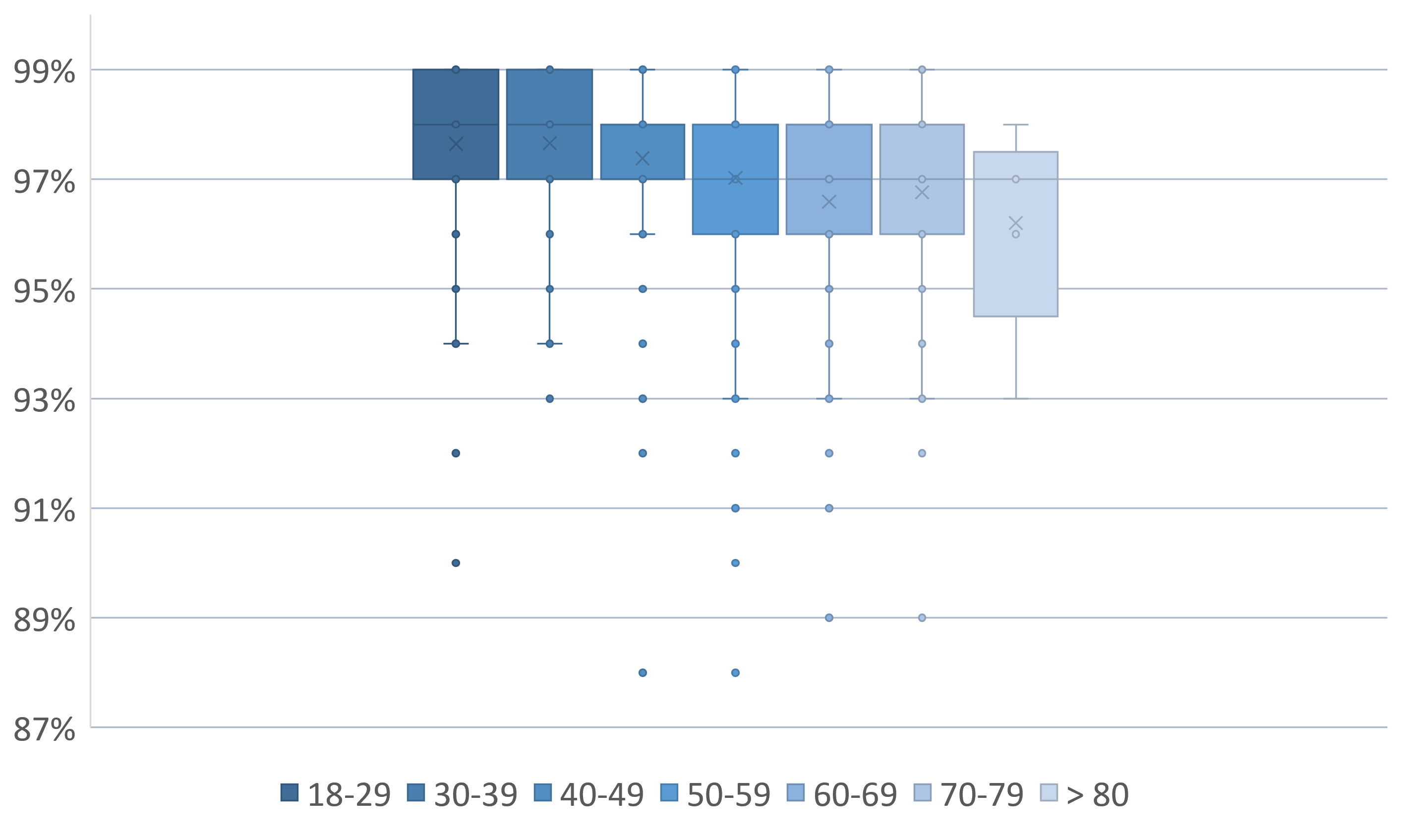}
          \label{fig:anxiety_a}
      \end{subfigure}
      \hfill
      \begin{subfigure}{0.49\textwidth}
        \includegraphics[width=\textwidth]{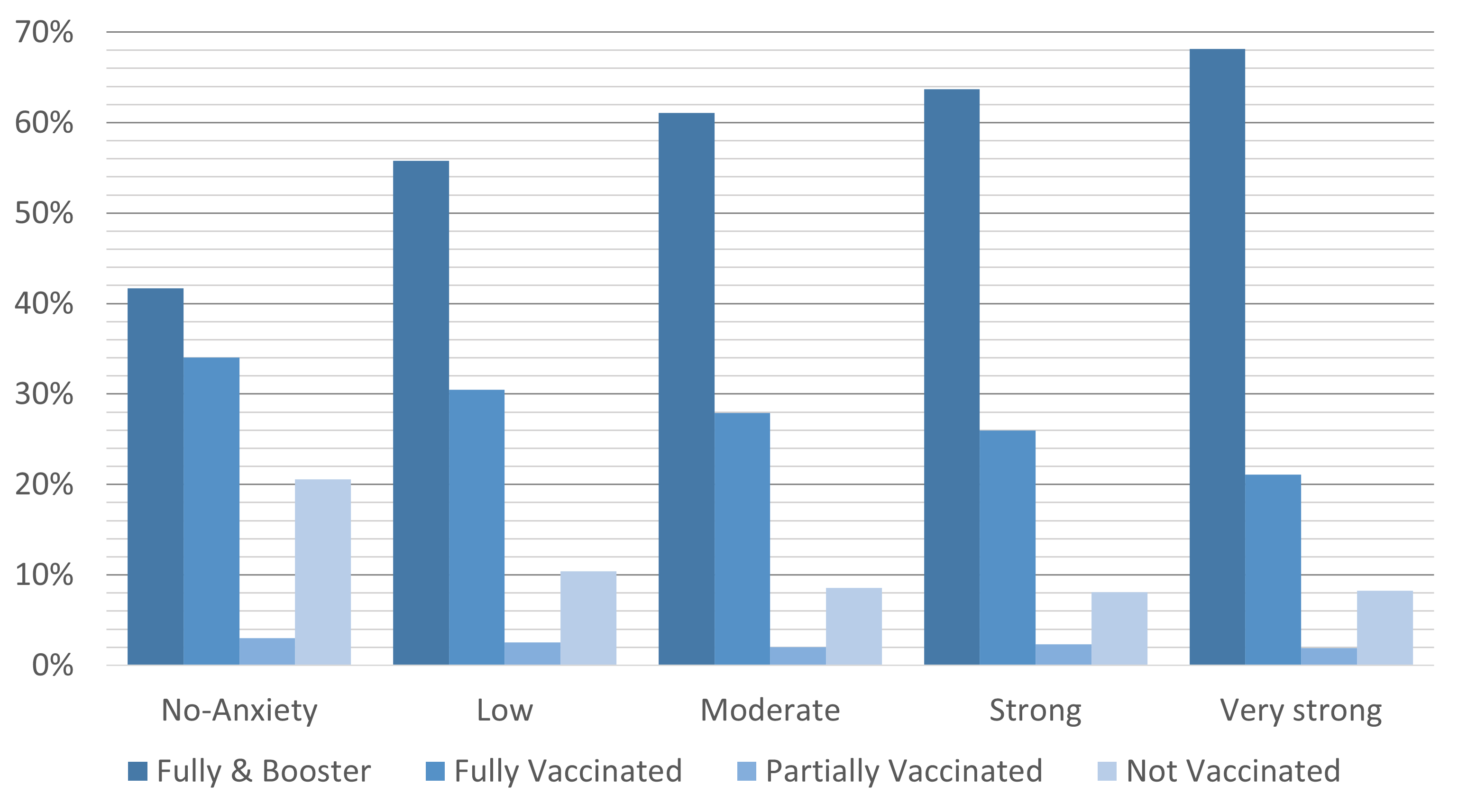}
          \label{fig:anxiety_b}
      \end{subfigure}
\caption{
\label{fig:anxiety}%
(a) Oxygen saturation vs age, (b) Level of COVID-19 related anxiety vs COVID-19 vaccination status.}
\end{figure}

The smarty4covid dataset also included vital signs (e.g oxygen saturation, beats per minute (BPM), diastolic/systolic pressure) as measured by means of relevant devices and self-reported COVID-19 related anxiety level. A box plot of the oxygen saturation for different age groups (Figure \ref{fig:anxiety}) presents oxygen saturation reduction against age progression. Figure \ref{fig:anxiety} depicts the vaccination status versus anxiety. Higher levels of anxiety presented higher percentage of users vaccinated with booster dose. 

\subsection*{Training AI models for classification of audio types} \label{classifier}
An AI-based model for classifying audio segments into cough, voice and breathing was developed utilizing the smarty4covid dataset in order to: (i) validate the quality of the smarty4covid dataset towards training an AI model with generalization capabilities, (ii) support the automated cleaning of crowd-sourced audio recordings, and (iii) be integrated in relevant crowd-sourcing platforms for detecting whether the submitted audio recording is valid and if needed to prompt the users to repeat the audio recording.
\subsubsection*{Architecture}
The classifier was based on the combined use of 2D Convolutional Neural Networks (CNN) that received as input the Mel spectrograms of audio segments of a specific duration (\emph{d}) and output the probability of detecting cough, breath, and voice. The frequency axis of the Mel spectrograms had size equal to 128, while the size of the time axis (\emph{d}) was a hyperparameter which was tuned through applying a grid search from 128 to 1024 corresponding to approximately 1 to 10 \textit{s} of audio, respectively. Each CNN consisted of $b$ stacked blocks containing $l$ convolutional layers followed by a 2x2 max pooling layer and a dropout layer with the dropout probability set to its default value equal to 0.5. The convolutional layers of each block featured $k$ 3x3 relu activated kernels and applied identical padding in order to ensure that the output of each layer had the same dimensions as its input. Finally, the output of the final convolutional layer was flattened and fed to a fully connected layer with 3 softmax activated neurons. A grid search was performed to realize optimal values of the hyperparameters $l$ (from 1 to 3), $k$ (from 64 to 128), and $b$ (from 3 to log2(d)). 

This architecture was inspired by the winning entry \footnote{https://ieee-dataport.org/analysis/ntuautn-ieee-covid-19-sensor-informatics-challenge} of the COVID-19 sensor informatics challenge hackathon \footnote{https://healthcaresummit.ieee.org/data-hackathon/}. It is relatively lightweight with ~300k trainable parameters, depending on the value of hyperparameter \textit{d}, and shallow, with at most seven convolutional layers, which makes it less prone to overfitting and speeds up the training procedure, when compared to larger neural networks. Another advantage of this architecture relies on combining CNNs featuring different time sizes $d$ of the considered segments resulting in a multi-scale modelling approach.   

\subsubsection*{Training procedure}
Prior to the training process, the audio signals were normalized, the leading and trailing silence were removed, and the Mel spectrograms were extracted. Each training instance was obtained by applying a randomized selection of a labeled (cough, voice, breathing) segment of a width $d$. The CNN's training procedure aimed at driving the optimization of the categorical cross entropy loss through the Adam algorithm \cite{kingma2014adam}. 
During inference, a sliding window of length $d$ and step $1$ was used to extract all (overlapping) segments of the audio signal, which were then fed to the trained CNN that estimated the probabilities of detecting cough, voice and breathing. Following this approach, the classification of an entire audio signal was also feasible, by combining (e.g averaging) the estimated probabilities over all extracted segments.

\subsubsection*{Results and external evaluation}
Aiming at exploring the impact of width ($d$) on the model's performance, a low (e.g. 128) and a high (e.g. 1024) value was applied resulting in two classifiers operating in short (1 \textit{s}) and long (10 \textit{s}) time scale, respectively. In order to evaluate the generalization capabilities of the classifiers, the COSWARA dataset served as external validation dataset since it includes all the three types of the considered audio recordings. Table \ref{tab:confusion} presents the confusion matrix of the obtained results. The long time scale classifier achieved a slightly better discrimination performance than the one obtained by applying the short time scale classifier (accuracy = $95.3\%$ vs $94\%$, c-statistic = $0.995$ vs $0.992$, macro F1 score = $0.953$ vs $0.941$). Leveraging upon the proposed architecture's flexibility, a multi-scale classifier was developed as an ensemble of the short and long time scale classifiers by applying a soft combination scheme (e.g. averaging) on the primary output probabilities. The obtained confusion matrix (Table \ref{tab:confusion}) indicated that the multiscale classifier had the highest sensitivity in detecting cough and breathing and the lowest one in detecting voice, yet the difference among the classifiers' performances was small. 

In order to justify the multiscale classifier's effectiveness, its performance was comparatively assessed with the one obtained by applying the COUGHVID classifier \cite{orlandic2021coughvid} on the coswara dataset, that was based on pretrained XGBoost and scaler. Table \ref{tab:coughvid_confusion} presents the confusion matrix when applying a probability decision threshold equal to $0.8$, which is the optimal threshold as mentioned by the creators of the COUGHVID model. The superiority of the multiscale classifier over the COUGHVID model was demonstrated through the evaluation metrics of accuracy ($95.4\%$ versus $83.6\%$), c-statistic ($0.995$ versus $0.888$) and macro F1 score ($0.954$ versus $0.81$).    

\subsection*{Conceptual edits on the smarty4covid OWL knowledge to produce counterfactual explanations} 
\begin{figure}[h!]
\centering
 \includegraphics[width=\textwidth]{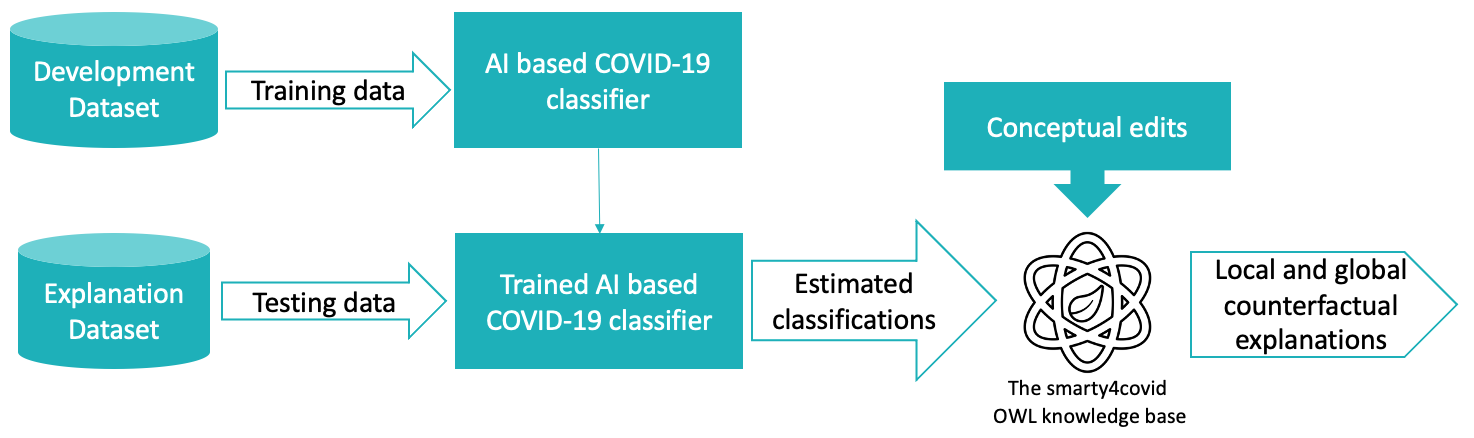}
\caption{Methodological framework towards producing counterfactual explanations of an AI based COVID-19 classifier utilizing the smarty4covid OWL knoweledge base}
\label{fig:counterfactual_explanations}%
\end{figure}

\begin{figure}[h!]
\centering
 \includegraphics[width=\textwidth]{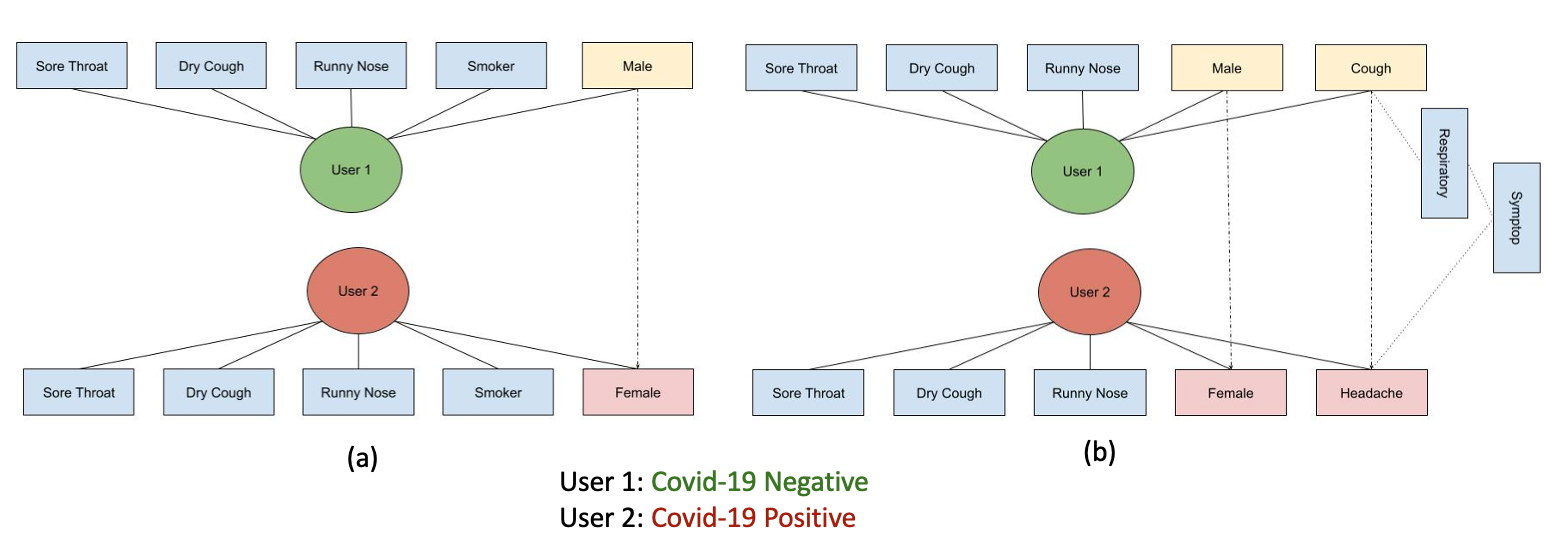}
\caption{Examples of applying conceptual edits on the smarty4covid OWL knowledge. User 1 and user 2 are the closest COVID-19 and non COVID-19 neighbours in the smarty4covid OWL knowledge. The minimal concept edits towards switching from positive to negative includes: (a) changing the gender from female to male, (b) changing the gender from female to male and the symptom from headache to cough}
\label{fig:conceptual_edits}%
\end{figure}

Taking into consideration the increased demand of transparent AI, a framework that leverages the high expressiveness of the smarty4covid OWL knowledge base is proposed towards identifying potential biases in the COVID-19 classification models and the datasets used for their development. The framework utilizes counterfactual explanations that can provide meaningful information by generating the most influencing factors affecting the model's output. As depicted in Figure \ref{fig:counterfactual_explanations}, it includes two different datasets: (i) the development dataset that is used to train an AI based classifier, and (ii) the explanation dataset that is used to test the trained AI based COVID-19 classifier. The trained AI based COVID-19 classifier is applied on the explanation dataset and the estimated classifications feed the smarty4covid OWL knowledge base by replacing the actual classifications (e.g. COVID-19, non COVID). The obtained modified smarty4covid OWL knowledge is subject to conceptual edits, which apply alterations on the concepts in order to identify the minimum changes that result in switching the estimated classification to a desired class. A thorough description of utilizing conceptual edits as counterfactual explanations is presented in \cite{filandrianos2022conceptual}. Figure \ref{fig:conceptual_edits} illustrates two examples of identifying the minimal conceptual edits in order for a positive COVID-19 user to become negative. The global counterfactual explanations are obtained by adding the minimal concepts edits over all users.        

\begin{figure}[h!]
\centering
 \includegraphics[width=0.8\textwidth]{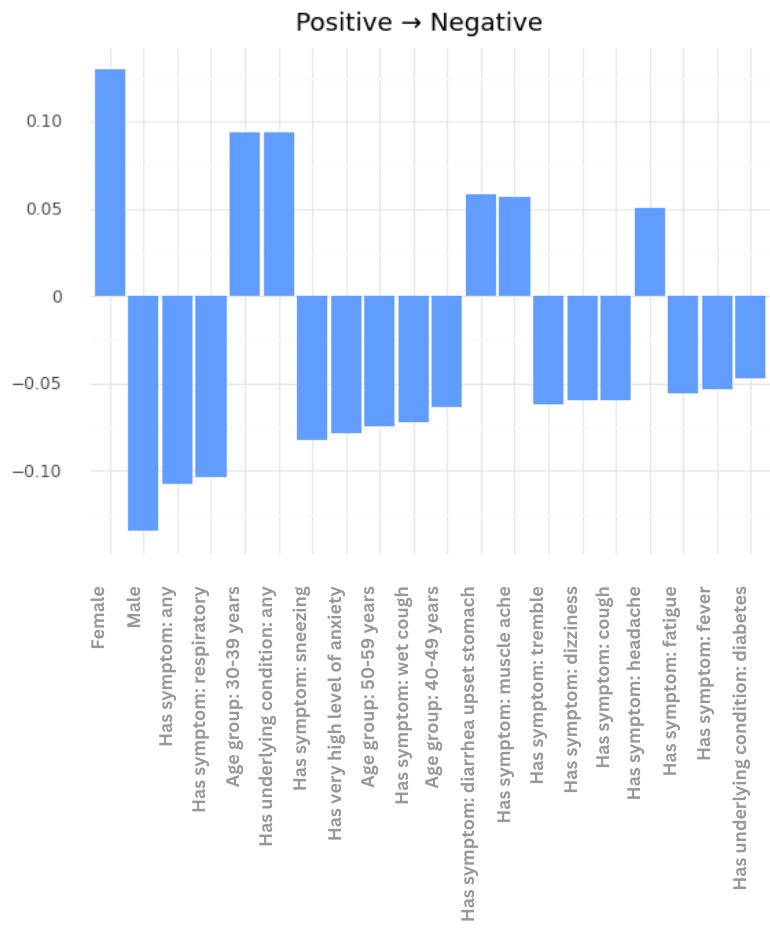}
\caption{Global counterfactual explanations taking into consideration the coswara dataset as development dataset and the smarty4covid dataset as explanation dataset.}

\label{fig:global_explanations}%
\end{figure}

In order to validate the aforementioned framework, a COVID-19 classifier was developed and potential biases were explored taking into consideration the Coswara dataset as development dataset and the smarty4covid dataset as explanation dataset. The COVID-19 classifier was based on ensembles of CNNs that received as input segments of the cough audio signal's mel spectrogram. The obtained global explanations are presented in Figure \ref{fig:global_explanations}. Gender was considered to be the most critical factor towards switching from positive for COVID-19 to negative. This was a bias that needed to be further explored whether it was a bias in the Coswara dataset or in the COVID-19 classifier. The application of some basic statistics on the Coswara dataset revealed that the COVID-19 prevalence in the male population was higher than the one in the female population. The same applied to the age of the Coswara's users as depicted in Figure \ref{fig:global_explanations}.  

\section*{Usage Notes}
A triplestore purpose-built database (e.g GraphDB \footnote{https://graphdb.ontotext.com}) is required in order to utilize the OWL knowledge base files while an ontology editor (e.g protege \footnote{http://protege.stanford.edu}) is needed to modify the underlying ontology. The smarty4covid OWL ontology can also be loaded as python object using the owlready2 \footnote{https://owlready2.readthedocs.io} and rdflib \footnote{https://rdflib.readthedocs.io} packages.

\section*{Code availability}
The audio classifier and the algorithm for extracting breathing features are available in a public repository \footnote{https://github.com/kinezodin/smarty4covid}. Furthermore, the repository includes the weights of the CNNs used by the classifier and a script for generating triples from the available data for the purpose of customizing the smarty4covid OWL knowledge base. 

\bibliography{sample}

\section*{Acknowledgements} 

This research was funded by the Hellenic Foundation for Research and Innovation-H.F.R.I within the framework of the H.R.F.I Science \& Society “Interventions to address the economic and social consequences of the COVID-19 pandemic” call. Grant number: 05020. 

\section*{Author contributions statement}
K.Z: study conception, design, and implementation, draft manuscript preparation, interpretation of results, funding acquisition; E.D, G.F, T.G: study implementation, data curation, data analysis and interpretation of results; V.G, A.S: data labeling; R.R, M.N: review and editing, interpretation of results; G.S: conceptualization, interpretation of results; K.N: study conception, interpretation of results, funding acquisition; supervision; All authors reviewed the manuscript.

\section*{Competing interests}
The authors declare no competing interests.

\section*{Figures \& Tables}

\begin{table}[!ht]
\centering
\begin{tabular}{@{}lcc@{}}
\toprule
Indicator          & Description & Normal range \\ \midrule
RR                             & Number of breaths per minute       &16-20                               \\
I/E Ratio                      & \textit{$T_{i}$}/{\textit{$T_{e}$}}    & 1:2 to 1:3 \\
FIT                            & \textit{$T_{i}$}/{\textit{$T_{tot}$}}    & 0.421 ± 0.033 \\ \bottomrule
\end{tabular}
\caption{Normal ranges of respiratory indicators}
\label{tab:normal_breath}
\end{table}

\begin{table}
\centering
\begin{tabular}{|l|l|l|l|l|}
\hline
    Field name & \multicolumn{2}{l|}{Description} & Type & Values \\
\hline
    participantid & \multicolumn{2}{l|}{Participant's Identification number} & String & UUID\\
\hline
    sex	& \multicolumn{2}{l|}{Participant's gender} & Int & \makecell[l]{0: Male, \\ 1: Female, \\ 2: Other} \\
\hline
age\_category & \multicolumn{2}{l|}{Age group} & Int & \makecell[l]{0: 18-29, \\ 1: 30-39, \\ 2: 40-49, \\ 3: 50-59, \\ 4: 60-69, \\ 5: 70-79, \\ 6: 80+}\\
\hline
bmi & \multicolumn{2}{l|}{Body mass index} & Float & [0,$\infty$)\\
\hline
asthma & \multirow{15}{*}{\begin{sideways}Underlying medical conditions\end{sideways}} & Asthma & Bool &  \\
respiratory\_deficiency & & Chronic respiratory failure / Emphysema & Bool &  \\
cystic\_fibrosis & & Cystic Fibrosis & Bool & \\
pneum\_other & & Other lung disease & Bool & \\
coronary\_disease	& &	Coronary artery disease	& Bool & \\
hypertension & &  Hypertension & Bool & \\
valve\_disease	& &	Valvular disease & Bool	& \\
heart\_attack	& &	History of heart attack &	Bool & 	\\
stroke	& &	History of stroke/ History of transient ischemic attack	& Bool &  \\
cardiovascular\_other & & Other cardiovascular disease & Bool & \\
diabetes &  &	Diabetes mellitus	& Bool & 	\\
kidney\_disease &  &	Chronic kidney disease / Chronic liver disease & Bool & \\
transplant & & Organ transplant	& Bool & 	\\
cancer &  &	Cancer in the last 5 years & Bool & \\
immunosuppression\_immunodeficiency	& &	Immunosuppression / Immunodeficiency &	Bool & \\
\hline
registration\_timestamp & \multicolumn{2}{l|}{Registration timestamp} & String & \\
\hline
\end{tabular}
\caption{Demographics and underlying conditions json file description}
\label{tab:demographics_json} 
\end{table}

\newcolumntype{C}[1]{>{\centering\let\newline\\\arraybackslash\hspace{0pt}}m{#1}}
\pagebreak
\begin{longtable}{|l|l|l|l|l|}
\hline
Field name & \multicolumn{2}{p{4cm}|}{Description} & Type & Values \\*
\hline
participantid & \multicolumn{2}{p{4cm}|}{Participant's identification number.} & String & UUID \\*
 \hline
 submissionid &	\multicolumn{2}{p{4cm}|}{Questionnaire's Identification number.}	& String & UUID\\*
 \hline
 covid\_status	& \multicolumn{2}{p{4cm}|}{Tested for COVID-19.}	& String & \makecell[l]{"positive": Positive, \\ "negative": Negative, \\ "no": Not tested}\\*
 \hline
 pcr\_test &	\multicolumn{2}{p{4cm}|}{Tested with PCR.} & Bool &\\*
 \hline
 rapid\_test & \multicolumn{2}{p{4cm}|}{Tested with a Rapid Antigen test.} &	Bool &\\
 \hline
 self\_test	& \multicolumn{2}{p{4cm}|}{Tested with a Rapid Antigen Self test.} &	Bool & \\
 \hline
 test\_last\_3\_days & \multicolumn{2}{p{4cm}|}{Tested in the last 3 days.} & Bool & \\
 \hline
 last\_negative\_test\_date	& \multicolumn{2}{p{4cm}|}{Date of the last negative test.} & String &	"yyyy-mm-dd" \\
 \hline
 first\_positive\_test\_date &	\multicolumn{2}{p{4cm}|}{Date of the first positive test.}	& String &	"yyyy-mm-dd"\\
 \hline
 vaccination\_status &	\multicolumn{2}{p{4cm}|}{COVID-19 vaccination status.}	& String & \makecell[l]{"no": No, \\ "partially": One of two shots, \\ "fully": Fully, \\ "booster1": Fully and Booster dose,\\ "booster2": Fully and two Booster doses}\\
 \hline
 latest\_vaccination\_date	& \multicolumn{2}{p{4cm}|}{Date of the last vaccination dose.} & String & "yyyy-mm-dd"\\
 \hline
 hospitalization & \multicolumn{2}{p{4cm}|}{Whether the user was hospitalised for COVID-19.} & String & \makecell[l]{ "0": "No", \\ "1": "I am currently hospitalized", \\ "2": "Yes, discharged a week ago", \\ "3": "Yes, discharged more than a month ago"}\\
 \hline
 exposure\_to\_someone\_with\_covid	& \multicolumn{2}{p{4cm}|}{Whether the user was exposed to a confirmed COVID-19 case.} & String & "No" / "Maybe" / "Yes" \\
 \hline
 travelled\_abroad	& \multicolumn{2}{p{4cm}|}{Whether the user has travelled abroad the last 14 days.} & String & \makecell[l]{"0": No,\\ "1": Yes} \\
 \hline
 submission\_timestamp &	\multicolumn{2}{p{4cm}|}{Timestamp when the submission was received} & String & \\
\hline
\caption{Main questionnaire json file description (Part 1/3: COVID-19 related information)}
\label{tab:main_questionnaire}
\end{longtable}
\pagebreak
\addtocounter{table}{-1}
\begin{longtable}{|l|l|l|l|l|}
\hline
Field name & \multicolumn{2}{p{4cm}|}{Description} & Type & Values \\*
\hline
sore\_throat & \multirow{16}{*}{\begin{sideways}Symptoms\end{sideways}} & Sore Throat & Bool & \\* 
 dry\_cough	&  &	Dry Cough	& Bool & \\*	
wet\_cough	& &	Productive Cough &	Bool &\\*
sputum & &	Sputum &	Bool &\\*
runny\_nose & &	 Nasal congestion &	Bool &\\*
breath\_discomfort	& &	Dyspnea &	Bool & \\*
has\_fever	& &	Fever &	Bool & \\*
tremble & &	Chills &	Bool	& \\*
fatigue	& &	Fatigue &	Bool	&\\*
headache & &	Headache &	Bool & \\*
dizziness & &	Dizziness/ confusion &	Bool & \\*
myalgias\_arthralgias	& &	Myalgias, arthralgias	& Bool & \\*
taste\_smell\_loss	& &	Loss of taste/smell	& Bool & \\*
diarrhea\_upset\_stomach	& &	Stomach upset/ Diarrhea	& Bool & \\*
sneezing & &		Sneezing	& Bool	& \\*
dry\_throat	& &	Dry Throat	& Bool	& \\*
\hline
oxymeter & \multirow{7}{*}{\begin{sideways}Vital Signs\end{sideways}} & Oximetry test & Bool & \\*
oxygenSaturation & &	Oxygen Saturation &	Int	& [60, 99]\\*
bpm	& &	Beats per minute (BPM)	& Int	& [30, 250]\\*
blood\_pressure\_meter	& &	Blood pressure test	& Bool	& \\*
systolic\_pressure	& &	Systolic Pressure &	Int &	[30, 260]\\*
diastolic\_pressure	& &	Diastolic Pressure & Int &	[30, 260] \\*
breath\_holding	& & Seconds of breath holding & Int &	[0, $\infty$)\\
\hline
leave\_bed & \multirow{6}{*}{\begin{sideways}Difficulty to\end{sideways}} & Leave Bed & Bool &\\*
leave\_home	& &	Leave Home &	Bool & \\*
prepare\_meal	& &	Prepare Meal & Bool & \\*
concentrate	& &	Concentrate	& Bool	& \\*
self\_care	& &	Self Care &	Bool & \\*
other\_difficulty & & Every day activites & Bool &\\*
\hline
\caption{Main questionnaire json file description (Part 2/3: Symptoms and vital signs)}
\end{longtable}
\pagebreak
\addtocounter{table}{-1}
\begin{longtable}{|l|l|l|l|l|}
\hline
Field name & \multicolumn{2}{C{4cm}|}{Description} & Type & Values \\*
\hline
smoking & \multirow{9}{*}{\begin{sideways}Smoking habits\end{sideways}}	& Smoking status	& String &	\makecell[l]{"nev": Never smoked,\\"ex": Ex-smoker,\\ "yes": Smoker}\\*
years\_of\_quitting\_smoking & & Years of quitting smoking	& Int	& [0, $\infty$)\\*
years\_of\_smoking	& &	Years of smoking &	Int & [0, $\infty$) \\*
no\_cigarettes	& &	Number of cigarettes per day &	String & \makecell[l]{"1u": less than 1,\\ "10u": 1-10,\\ "20u": 11-20,\\"20o": more than 20}  \\*
vaping	& &	Vaping	& String & \makecell[l]{"0": No, \\"1": Yes}  \\*
\hline
anxiety	& \multicolumn{2}{p{4cm}|}{Level of anxiety about the pandemic} &	String & \makecell[l]{"0": None,\\"1": Low,\\"2": Moderate,\\"3": High,\\"4": Very High}\\*
\hline
working &	\multicolumn{2}{p{4cm}|}{Working Status} &	String	& \makecell[l]{"home": Working from home,\\*
"hospital": Working in hospital,\\*
"store": Working in an essential \\*goods store (pharmacy, supermarket),\\*
"social": Working in a service with increased \\* contact with the general public,\\*
"no": Not working}\\*
\hline
\caption{Main questionnaire json file description (Part 3/3: Smoking habits, anxiety level, and working status)}
\end{longtable}

\begin{table}[ht]
\centering
\begin{tabular}{|p{3cm}|p{8cm}|p{2cm}|p{3cm}|}
\hline
Field name	& Description &	Type &	Values \\
\hline
RR	& Estimated respiratory rate &	Float &	[0, $\infty$)\\
\hline
I\_E\_ratio	& Estimated Inhalation/Exhalation Ratio	& Float &	[0, $\infty$)\\
\hline
FIT &	Estimated Fractional Inspiration Time &	Float &	[0, $\infty$)\\
\hline
annotated\_inhalation	& Manual annotations of inhalation parts. Contains a list of (start, end) pairs indicating the start and the end time (in s) of an inhalation. &	List of tuples	& [(start1, end1), …, (startN, endN)]\\
\hline
annotated\_exhalation &	Manual annotations of exhalation parts. Contains a list of (start, end) pairs indicating the start and the end time (in s) of an exhalation. &	List of tuples	& [(start1, end1), …, (startN, endN)]\\
\hline
\end{tabular}
\caption{Breathing features json file description}
\label{tab:breathing_features}
\end{table}

\begin{table}[ht]
\begin{tabular}{|p{7cm}|p{0.3cm}|p{5cm}|p{0.7cm}|p{2.5cm}|}
\hline
Field name & \multicolumn{2}{C{5.7cm}|}{Description} & Type & Values\\
\hline
breath\_depth & \multicolumn{2}{C{5.7cm}|}{Depth of breathing} & String & “Can breathe deeply enough”/“Cannot breathe deeply enough”\\
\hline
dyspnea\_shortness\_of\_breath & \multirow{11}{*}{\begin{sideways}Audible abnormalities \end{sideways}} &	Dyspnea, Shortness of breath &	Bool & \\
stridor	 & &	Stridor &	Bool	&\\
inspiratory\_stridor	& &	Inspiratory stridor	& Bool	&\\
expiratory\_stridor	& &	Expiratory stridor &	Bool	& \\
wheezing	& &	Wheezing	& Bool	& \\
respiratory\_crackles	& &	Respiratory crackles &	Bool &\\
prolonged\_expiration	& &	Prolonged expiration &	Bool &\\
other\_and\_unspecified\_abnormalities\_of\_breathing	& &	Other and unspecified abnormalities of breathing & Bool &\\
audible\_choking	& &	Audible choking	& Bool	& \\
audible\_nasal\_congestion	& &	Audible nasal congestion &	Bool	& \\
no\_audible\_abnormalities	& &	No audible abnormalities	& Bool	& \\
\hline
\end{tabular}
\caption{Experts’ breath annotation json file description}
\label{tab:experts_breathing}
\end{table}

\begin{table}[ht]
\begin{tabular}{|p{4cm}|p{0.8cm}|p{3.5cm}|p{0.7cm}|p{7cm}|}
\hline
Field name & \multicolumn{2}{C{4.5cm}|}{Description} & Type & Values\\
\hline
sex	& \multicolumn{2}{C{4cm}|}{Expert’s opinion on the users’ gender based on cough recording}	& String &	“Male”, “Female”,“Can’t tell”\\
\hline
patient\_has & \multirow{2}{*}{\begin{sideways} Expert’s impression\end{sideways}} &	“I think this patient has:” &	String	& \makecell[l]{“An upper respiratory tract infection”,\\ “A lower respiratory tract infection”,\\“Obstructive lung disease (Asthma, COPD, ...)”,\\“Nothing (healthy cough)”} \\
cough\_is	& &	“The cough is probably:” &	String	& \makecell[l]{“Pseudo cough/Healthy cough”,\\ “Mild (from a sick person)”,\\ “Severe (from a sick person)”,\\“Can’t tell”}\\
\hline
productive & \multirow{7}{*}{\begin{sideways}Type of cough\end{sideways}} &	Productive & Bool, & \\
dry	& &	Dry &	Bool	&\\
barking\_cough	& &	Barking cough &	Bool	&\\
hacking\_cough	& &	Hacking cough &	Bool	&\\
croupy\_cough	& &	Croupy cough &	Bool &\\
other\_specified\_cough	& &	Other specified cough &	Bool & \\
can't\_tell	& &	Can't tell &	Bool	 & \\ 
\hline
audible\_dyspnea &	\multirow{6}{*}{\begin{sideways}\makecell[c]{Audible \\abnormalities}\end{sideways}} & Audible dyspnea &	Bool & \\
audible\_wheezing	& &	Audible wheezing	& Bool	& \\
audible\_stridor	& &	Audible stridor &	Bool	& \\
audible\_choking	& &	Audible choking &	Bool	&\\
audible\_nasal\_congestion & &	Audible nasal congestion &	Bool &\\
nothing\_specific	& &	Nothing specific &	Bool &\\	
\hline
\end{tabular}
\caption{Experts’ cough annotation json file description}
\label{tab:experts_cough}
\end{table}

\begin{table}[ht]
\begin{tabular}{|p{4cm}|p{0.8cm}|p{3.5cm}|p{0.7cm}|p{4cm}|}
\hline
Field name & \multicolumn{2}{C{6cm}|}{Description} & Type & Values\\
\hline
completion	& \multicolumn{2}{C{6cm}|}{Expert’s answer whether the user completed the phrase} &	String &	“Yes”, “No” \\
\hline
hoarseness & \multicolumn{2}{C{4.5cm}|}{Hoarseness of voice} & String	& “Yes”, “No”\\
\hline
volume &	\multicolumn{2}{C{4.5cm}|}{Vocal intensity} & String & “Low (whisper)”, “Normal”\\
\hline
audible\_dyspnea &	\multirow{6}{*}{\begin{sideways}\makecell[c]{Audible \\abnormalities}\end{sideways}}	& Audible dyspnea &	Bool & \\
audible\_wheezing & &		Audible wheezing	& Bool	&\\
audible\_stridor & &		Audible stridor &	Bool	&\\
audible\_choking & &		Audible choking	& Bool &\\
audible\_nasal\_congestion	& &	Audible nasal congestion	& Bool	&\\
no\_audible\_abnormalities	& &	No audible abnormalities	& Bool	&\\
\hline
\end{tabular}
\caption{Experts’ voice annotation json file description}
\label{tab:experts_voice}
\end{table}

\begin{table}[h!]
\centering
\begin{tabular}{|c|c|c|c|c|c|c|c|c|c|}
\hline
\multirow{2}{*}{\diagbox{Detected}{Actual}} & \multicolumn{3}{c|}{Short scale (($d=128$))} & \multicolumn{3}{c|}{Large scale (($d=1024$))} & \multicolumn{3}{c|}{Multiscale scale} \\ \cline{2-10}
 & Cough & Breath & Speech  & Cough & Breath & Speech  & Cough & Breath & Speech\\ \hline
Cough  & 944   & 75     & 21  & 948   & 64     & 17 & 949   & 60     & 24   \\ \hline
Breath & 15    & 845    & 9  & 7     & 873    & 10  & 11    & 881    & 10    \\ \hline
Speech & 6     & 45     & 935  & 10    & 28     & 938 & 5     & 24     & 931 \\ \hline
\end{tabular}
\caption{Confusion matrices of the short, long and multi- time scale classifiers when evaluated on the coswara dataset}
\label{tab:confusion}
\end{table}

\begin{table}[h!]
\centering
\begin{tabular}{|l|l|l|l|}
\hline
\diagbox{Detected}{Actual} & Cough & Breath & Speech \\ \hline
Cough    & 677   & 163    & 24     \\ \hline
No Cough & 288   & 802    & 941    \\ \hline
\end{tabular}
\caption{Confusion matrix of the cough-detection model provided by COUGHVID when evaluated on the coswara dataset.}
\label{tab:coughvid_confusion}
\end{table}

\end{document}